\documentclass[aps,prl,twocolumn,superscriptaddress]{revtex4-2}

\usepackage{comment}
\usepackage{graphicx}
\usepackage{multirow}
\usepackage{amsmath}
\usepackage{longtable}
\usepackage{xspace}
\usepackage[usenames,dvipsnames]{xcolor}
\usepackage{soul}
\usepackage{hyperref}
\usepackage[normalem]{ulem}
\usepackage{comment}

\newcommand*{\msr}{$\mu$SR\xspace}

\begin{document}

%Title of paper
\title{Magnetostriction-driven muon localisation in an antiferromagnetic oxide}

\author{Pietro Bonf\`a}
\email[]{pietro.bonfa@unipr.it}
\affiliation{Dipartimento di Scienze Matematiche, Fisiche e Informatiche, Universit\'a di Parma, I-43124 Parma, Italy}

\author{Ifeanyi John Onuorah}
\affiliation{Dipartimento di Scienze Matematiche, Fisiche e Informatiche, Universit\'a di Parma, I-43124 Parma, Italy}

\author{Franz Lang}
\affiliation{ISIS Neutron and Muon Source, STFC Rutherford Appleton Laboratory, Chilton, Didcot OX11 0QX, United Kingdom}

\author{Iurii Timrov}
\affiliation{Theory and Simulation of Materials (THEOS), and National Centre for Computational Design and Discovery of Novel Materials (MARVEL), \'Ecole Polytechnique F\'ed\'erale de Lausanne, 1015 Lausanne, Switzerland}

\author{Lorenzo Monacelli}
\affiliation{Theory and Simulation of Materials (THEOS), and National Centre for Computational Design and Discovery of Novel Materials (MARVEL), \'Ecole Polytechnique F\'ed\'erale de Lausanne, 1015 Lausanne, Switzerland}

\author{Chennan Wang}
\affiliation{Laboratory for Muon Spin Spectroscopy, Paul Scherrer Institute, CH-5232 Villigen, Switzerland}

\author{Xiao Sun}
\affiliation{Jülich Centre for Neutron Science JCNS-2 and Peter Grünberg Institute
PGI-4, JARA-FIT, Forschungszentrum Jülich GmbH, 52425 Jülich, Germany}
\altaffiliation{Present address of Xiao Sun: Deutsches Elektronen-Synchrotron DESY, Photon Science, 
22607 Hamburg, Germany.}

\author{Oleg Petracic}
\affiliation{Jülich Centre for Neutron Science JCNS-2 and Peter Grünberg Institute
PGI-4, JARA-FIT, Forschungszentrum Jülich GmbH, 52425 Jülich, Germany}

\author{Giovanni Pizzi}
\affiliation{Laboratory for Materials Simulations (LMS), Paul Scherrer Institut (PSI), CH-5232 Villigen PSI, Switzerland}

\author{Nicola Marzari}
\affiliation{Theory and Simulation of Materials (THEOS), and National Centre for Computational Design and Discovery of Novel Materials (MARVEL), \'Ecole Polytechnique F\'ed\'erale de Lausanne, 1015 Lausanne, Switzerland}
\affiliation{Laboratory for Materials Simulations (LMS), Paul Scherrer Institut (PSI), CH-5232 Villigen PSI, Switzerland}
\author{Stephen J. Blundell}
\affiliation{Department of Physics, University of Oxford, Clarendon Laboratory, Oxford OX1 3PU, United Kingdom}

\author{Roberto De Renzi}
\affiliation{Dipartimento di Scienze Matematiche, Fisiche e Informatiche, Universit\'a di Parma, I-43124 Parma, Italy}

\date{\today}

\begin{abstract}
Magnetostriction results from the coupling between magnetic and elastic degrees of freedom.  Though it is associated with a relatively small energy, we show that it plays an important role in determining the site of an implanted muon, so that the energetically favorable site can switch on crossing a magnetic phase transition.  This surprising effect is demonstrated in the cubic rock salt antiferromagnet MnO which undergoes a magnetostriction-driven rhombohedral distortion at the N\'eel temperature $T_{\rm N}=118$~K.  Above $T_{\rm N}$, the muon becomes delocalized around a network of equivalent sites, but below $T_{\rm N}$ the distortion lifts the degeneracy between these equivalent sites.  Our first-principles simulations based on Hubbard-corrected density-functional theory and molecular dynamics are consistent with the experimental data and help to resolve a long-standing puzzle regarding muon data on MnO, as well as having wider applicability to other magnetic oxides.
\end{abstract}

\maketitle

The coupling between the magnetization and the lattice can result in a deformation, called magnetostriction. This magnetostructural interaction is rather weak and, for example in insulating magnets containing transition metal ions, is dwarfed by the much larger magnetic superexchange interaction between localized magnetic moments.
A commonly used technique to study such magnetic materials is
muon-spin spectroscopy (\msr \cite{blundell2022}), in which a spin-polarized positive muon is implanted in a sample primarily under the effect of electrostatic forces (therefore preserving its spin polarization while losing kinetic energy).
For this reason, it has not been expected that the much smaller magnetostructural couplings should play any role in determining the experimental signal. In this Letter, we demonstrate the surprising fact that, below a magnetic phase transition, exchange-driven effects can drastically change the nature of the muon state, so that the muon switches its energetically-favorable position as the sample is cooled through $T_{\rm N}$. Our study is focused on the prototypical antiferromagnet manganese oxide (MnO), but we describe an approach that has wider applicability in other magnetic oxides.

Magnetic order in MnO was identified over 65 years ago
\cite{rothMagneticStructuresMnO1958} and later
progressively refined with numerous studies \cite{PhysRevB.38.11901,PhysRevB.27.6964,PhysRevLett.96.047209,gazzara_middleton_1965, PhysRevB.28.6542, yamamotoSpinArrangementsMagnetic1972, PhysRev.142.287, 10.1143/JPSJ.15.466}.
%In addition to fundamental physics research, MnO is also found in a variety of applications, including catalysis, electronics, digital storage, spintronics, and in size-engineering through nanoparticle synthesis \cite{adma.201905823, acsomega.0c03455, PhysRevB.95.134427}. 
For temperatures above 
$T_{N}=118$~K,
MnO has the cubic rock salt structure ($Fm\bar{3}m$).
Below $T_{N}$ a magnetic transition to antiferromagnetic ordering (type-II) occurs, with Mn moments aligned ferromagnetically along (111) planes (likely along the [11$\bar{2}$] direction \cite{PhysRevLett.96.047209}), and anti-parallel between adjacent planes, which induces a small  distortion.  This magnetostrictive effect, associated with a deviation
of $\sim 0.6^\circ$ from the $90^\circ$ angle in the cubic structure \cite{PhysRevB.1.236, balagurovMagnetostructuralPhaseTransitions2016}, results in a rhombohedral distortion that is further refined by neutron scattering experiments into the monoclinic $C2/c$
symmetry of the magnetically ordered phase. An additional modulation of the atomic positions further reducing
the symmetry to $C2$ has also been suggested \cite{PhysRevLett.96.047209}.

Previous experiments on MnO using \msr~\cite{blundell2022}  
show that
($i$)~in zero applied field (ZF),
a single precession frequency is observed
for all $T<T_{N}$ \cite{Uemura1981b,Uemura1984}, which, up to now has been interpreted  to imply a highly symmetric muon site, identified as the $(\frac{1}{4},\frac{1}{4},\frac{1}{4})$ interstitial (8c) of the 
conventional cubic cell;
($ii$)~transverse field (TF) measurements for $T>T_{N}$ reveal a large negative Knight shift which is not proportional to the susceptibility \cite{uemuraSpinRelaxationPositive1979, Uemura1981a}, but shows an unusual  \emph{time dependence}, so that
the estimate of the Knight shift depends on the time interval used to fit the experimental asymmetry \cite{Ishida1984}.
This latter observation can be rationalized \cite{uemuraSpinRelaxationPositive1979} by assuming that the muon exhibits a thermally activated diffusion between sites and ends up close to a Mn vacancy \cite{aschauerEffectEpitaxialStrain2015, logsdailHybridDFTModelingLattice2019}, with the analysis of the  
time dependence of the Knight shift
yielding an activation energy for this process of about $800\text{ K}$ \cite{hayanoLongitudinalSpinRelaxation1978, uemuraSpinRelaxationPositive1979}.
An additional transition between 400~K and 600~K has been detected in the exponential depolarization rate of the ZF signal \cite{lidstromParamagneticFluctuationsMnO2000a} and may be related to a change in the orbital occupation of Mn-$3d$
orbitals \cite{PhysRevLett.62.478, PhysRevB.42.11895}, though this interpretation has not been supported by a quantitative prediction.
Moreover, the suggested muon site at the interstitial $(\frac{1}{4},\frac{1}{4},\frac{1}{4})$ position would place the muon surprisingly far from the electronegative oxygen atoms, but displacing it from this symmetrical position was assumed to lead to more than one local field at the muon site below $T_N$, in contrast with experiment.  The single precession frequency for $T<T_{N}$ is confirmed by our new results collected on a single crystal of MnO (characterized in Ref.~\cite{Sun_2017}). The effect is demonstrated in both the raw muon asymmetry [Fig.~\ref{fig:experiment}(a)] and in its Fourier transform [Fig.~\ref{fig:experiment}(b)]; see the Supplemental Information (SI) \footnote{See Supplemental Material at [URL will be inserted by publisher] for the details of the experimental setup, the computational methods, the details of the simulations and the analysis of experimental and theoretical data. It also includes Refs.~\cite{10.1063/5.0005077,Baroni:2001,BENNETT2019109137,Bonfa2016,Campo:2010,chmiela2023,Cococcioni:2005,DALCORSO2014337,Dudarev:1998,GARRITY2014446,giannozzi2020quantum,Giannozzi_2009,Giannozzi_2017,gipaw-website,Gorni:2018,He2018,Hohenberg:1964,Kohn:1965,Kucukbenli:2014,Kulik:2011,Lowdin:1950,MaterialsCloud,MaterialsCloudArchive2023,Mayer:2002,PhysRev.182.280,PhysRevA.36.958,PhysRevB.1.236,PhysRevB.23.5048,PhysRevB.27.6964,PhysRevB.38.11901,PhysRevB.7.5212,PhysRevB.97.174414,PhysRevLett.100.136406,PhysRevMaterials.3.073804,prandini2018precision,PRATT2000710,Timrov:2018,Timrov:2020b,Timrov:2021,Timrov:2022,Uemura1981a,Wang:2016,Tablero:2008,Amadon:2008,Nawa:2018}.} 
for further details.

\begin{figure}
    \includegraphics[width=\columnwidth]{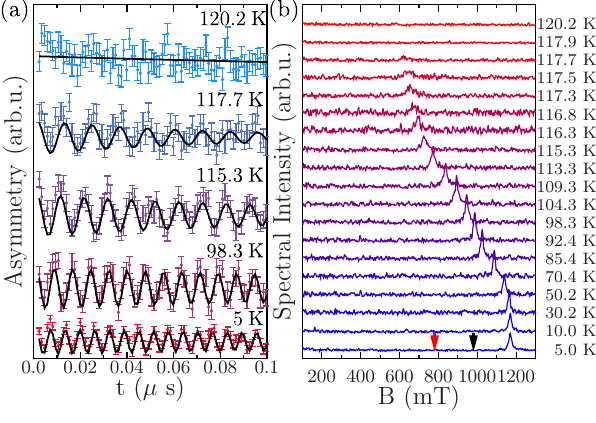}
    \caption{(a)~Muon asymmetry $A(t)$ of single crystal MnO at selected temperatures. The black line is a fit to the function $A(t)=A_b + A_o \cos(\gamma_{\mu} B_\mu t)\exp(-\lambda t)$ where $B_\mu$ is the local field at the muon site, $\gamma_\mu$ is the muon gyromagnetic ratio and the baseline ($A_b$) and oscillating ($A_o$) amplitudes depend on the setup of the experiment. (b)~Fourier transform of $A(t)$ at various temperatures. The red and black arrows are the prediction of the local field at the muon site obtained from DFT considering \emph{ab initio} or experimental contact field, respectively, while the dipolar part is computed assuming $4.9\;\mu_\text{B}$ per Mn atom. Data for various temperatures are displaced vertically on both panels for clarity. \label{fig:experiment}}
\end{figure}

\begin{table*}

\begin{ruledtabular}
\begin{tabular}{*{8}{c}}
Lattice       & Site & Position & $E-E_{0}$ & $\mathbf{B}_{\text{dip}}$ & $\mathbf{B}_{\text{cont}}$ & $|\mathbf{B}_{\text{dip}}|$ & $|\mathbf{B}_{\text{dip}}+\mathbf{B}_{\text{cont}}|$ \\
\hline
\multirow{2}{*}{Rhombohedral} & \multirow{2}{*}{1}  & (0.19, 0.19, 0.19) & \multirow{2}{*}{0}  & (-0.47, -0.47, 0.94) & ( 0.1, 0.1, -0.2) &  \multirow{2}{*}{1.13} &  \multirow{2}{*}{0.8}  \\
                       &    & (0.31, 0.31, 0.31) &   & (0.47, 0.47, -0.94)  & (-0.1, -0.1, 0.2) &   &  \\
                       \hline
\multirow{8}{*}{Cubic} & \multirow{2}{*}{1}    & (0.19, 0.19, 0.19) & \multirow{2}{*}{32} & (-0.42,  -0.42,  0.84)  & (0.1, 0.1,  -0.3) & \multirow{2}{*}{1.12} & \multirow{2}{*}{0.8} \\
                       &      & (0.31, 0.31, 0.31) &   & (0.42,  0.42, -0.84)  & (-0.1, -0.1,  0.3) &   &   \\ \cline{2-8}
                       & \multirow{4}{*}{2}    & (0.31, 0.19, 0.31) & \multirow{4}{*}{0}  &  (0.68, -0.30, -0.81)  & \multirow{4}{*}{$< 0.1$} & \multirow{4}{*}{1.13}  & \multirow{4}{*}{1.1} \\
                       &      & (0.31, 0.19, 0.19) &    &  (0.30, -0.68, 0.81) &   &   & \\
                       &      & (0.19, 0.31, 0.19) &    &  (-0.68, 0.30, 0.81)   &   &   & \\
                       &      & (0.19, 0.31, 0.31) &    &  (-0.30, 0.68, -0.81) &   &   & \\ \cline{2-8}
                       & \multirow{2}{*}{3}    & (0.31, 0.31, 0.19) & \multirow{2}{*}{0}   & (0.13, 0.13, 0.58)  & \multirow{2}{*}{$< 0.1$}   & \multirow{2}{*}{0.57} & \multirow{2}{*}{0.6}\\
                       &      & (0.19, 0.19, 0.31) &    & (-0.13, -0.13, -0.58)  &   &   & 
\end{tabular}
\end{ruledtabular}

\caption{Results of \emph{ab initio} analysis of muon sites. 
Energies are in meV relative to the lowest energy site and local fields are in Tesla.
Sites are labeled according to Fig.~\ref{fig:sites} and their position is in crystal coordinates with respect to the  $2 \times 2 \times 2$ conventional cubic/rhombohedral supercell where Mn is at the origin.
$\mathbf{B}_{\text{dip}}$ is the dipolar field at the muon sites computed in the locally distorted lattice.
$\mathbf{B}_{\text{cont}}$ is the contact field obtained from DFT simulations. The last two columns are the absolute values of the dipolar contribution and the
total field at the muon site in the AFM phase.
\label{tab:muons}}

\end{table*}

To interpret the experimental data, we performed density-functional theory (DFT)~\cite{Hohenberg:1964, Kohn:1965} simulations using Hubbard corrections~\cite{anisimov:1991, Dudarev:1998}, in particular using extended Hubbard functionals~\cite{Campo:2010}.
The electronic structure of MnO is described using the PBEsol functional~\cite{PhysRevLett.100.136406}, while self-interaction errors are alleviated using on-site $U=4.84$~eV and inter-site $V=0.36$~eV Hubbard parameters for Mn($3d$) and Mn($3d$)--O($2p$), respectively, computed self-consistently using density-functional perturbation theory \cite{Timrov:2018, Timrov:2021}. The additional inter-site parameter $V=0.50$~eV is obtained for Mn($3d$)--H($1s$) when considering the muon.
For \emph{ab initio} molecular dynamics simulations, used to train machine-learned force fields, and for nudged elastic band simulations (see SI \cite{Note1}), we used the PZ-LDA functional\cite{PhysRevB.23.5048} and an averaged on-site Hubbard $U = 5$~eV.

The rhombohedral distortion of the antiferromagnetic (AFM) phase \footnote{DFT simulations do not identify the small monoclinic distortion for MnO, see for example Ref.~\cite{schronCrystallineMagneticAnisotropy2012}}, predicted by DFT as reported in many previous studies 
\cite{PhysRevB.38.11901,PhysRevB.68.140406, PhysRevB.94.165151, paskStructuralElectronicMagnetic2001, balagurovMagnetostructuralPhaseTransitions2016, schronCrystallineMagneticAnisotropy2012, franchiniDensityFunctionalTheory2005},
is well reproduced
and the PBEsol+$U$+$V$ unit cell volume of 22.14~\AA$^{3}$ (21.89~\AA$^{3}$ within LDA$+U$) deviates from the experiment only by $\sim 2$\%.
The addition of a gradient correction in the exchange-correlation term and the inter-site contribution in the Hubbard correction alter the position found for the muon by a small margin ($<0.1$~\AA), slightly alter the contact term (by $\sim$15\%), and enhance embedding site energy differences (by $\sim$40\%).

\begin{figure}
    \includegraphics[width=\columnwidth]{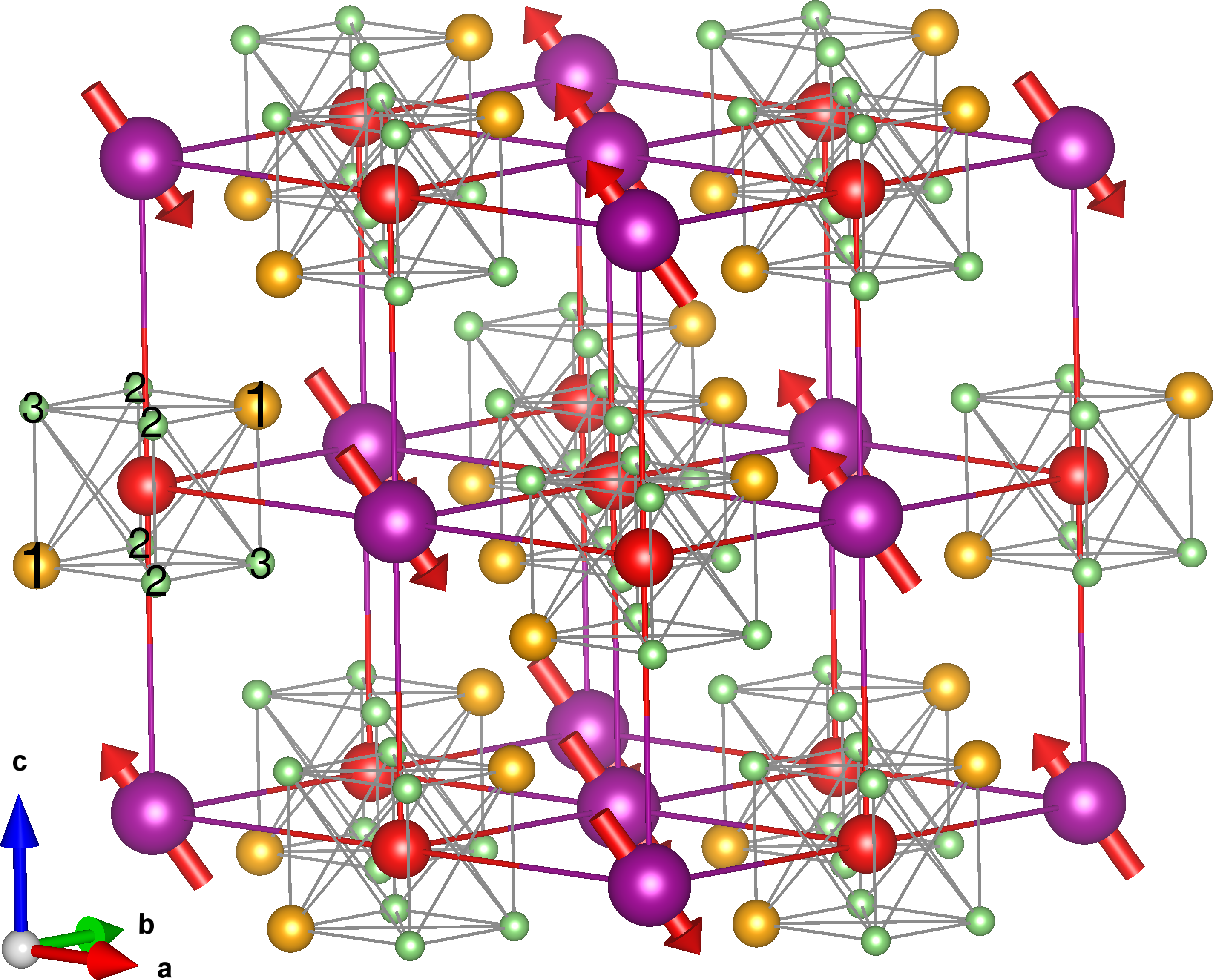}
    \caption{Muon sites in MnO. The arrows on Mn (purple) spheres show the magnetic order; oxygen is shown as red spheres.
    The muon sites are shown by orange and green spheres (to distinguish symmetrically inequivalent sites 2c and 6h in the rhombohedral cell) and labeled 1, 2 and 3 to identify muon sites with the same $|\mathbf{B}_{\mu}|$ in the 
    AFM phase. \label{fig:sites}}
\end{figure}

From an analysis of the muon embedding sites two important points emerge, providing a novel interpretation of the experimental findings:
$(i)$~$\mu^+$ does not stop in the $(\frac{1}{4},\frac{1}{4},\frac{1}{4})$ position and the stable locations are instead closer to the oxygen atoms, and $(ii)$~different equilibrium positions are observed in the low-temperature rhombohedrally distorted structure and in the high-temperature cubic lattice (with Wyckoff site symmetries 2c and 6h in the rhombohedral [$R\bar{3}m$] and 32f in the cubic [$Fm\bar{3}m$] phases).
The equilibrium positions (Table~\ref{tab:muons}) for both structures are shown in Fig.~\ref{fig:sites}, and, collectively, they form a cube around each oxygen atom.
Two colors are used to distinguish the geometrically inequivalent sites in the
rhombohedral cell, while the numbers identify sites with different absolute values of the local field in the AFM phase. In the rhombohedral phase there are only sites of kind 1 (orange spheres) since the higher energy sites (green spheres) are unstable owing to the absence of any barrier separating them from the lowest energy sites (see SI~\cite{Note1}).
The direct consequence of this observation is that these interstitial muon sites close to oxygen atoms produce a single precession frequency in the AFM phase, compatible with experimental observations without requiring the $\mu^+$ to stop far away from the O atoms in $(\frac{1}{4},\frac{1}{4},\frac{1}{4})$ and thus solving one of the points of the puzzle. 
This frequency can be estimated from 
$\mathbf{B}_{\mu} = \mathbf{B}_{\text{dip}} + \mathbf{B}_{\text{cont}}$, which is the sum of the dipolar and Fermi contact contributions.
The first term, due to the dipolar interaction of the muon with the distant $3d$ spin-polarized electrons, is computed assuming (classical) magnetic dipoles $\mathbf{m}$ at the Mn atomic positions. We set $\mathbf{m}=4.9\,\mu_{\mathrm B}$ following the recent experimental estimates \cite{PhysRevB.27.6964, mnmoment,BONFANTE1972553, PhysRevB.83.214418}
and we consider the displacement due to the presence of the interstitial positive muon \cite{HUDDART2022108488}. 
The contact term is instead evaluated from DFT \footnote{We performed collinear DFT simulations,
therefore the direction of $\mathbf{S}_{e}$ cannot be established. See supplemental information for more details on the approximations concerning the evaluation of the Fermi contact field.} (reported as $B_{\text{cont}}$), or obtained experimentally (see SI \cite{Note1}).

The estimated local fields at the muon sites are summarized in Table~\ref{tab:muons}.
The local field at the muon site appears to be slightly underestimated, with a predicted value of $B_{\mu}=0.8$~T instead of the 1.17~T observed experimentally for $T \to 0$~K, as shown by the red arrow in Fig.~\ref{fig:experiment}. 
Notably, the (negative) contact hyperfine coupling estimated from Hubbard-corrected DFT is slightly larger than the one obtained from TF experiments~\cite{Uemura1981a}.
This discrepancy is not surprising \cite{PhysRevB.97.174414}
and the estimate may possibly be improved by taking into account anharmonic effects \cite{PhysRevMaterials.3.073804}.
The smaller contact field contribution obtained in Ref.~\cite{Uemura1981a} improves the experimental agreement, as shown by the black arrow in Fig.~\ref{fig:experiment}.
The additional monoclinic distortion of the structure \cite{PhysRevLett.96.047209} does not alter this picture, but the further modulation of atomic positions discussed in Ref.~\cite{PhysRevLett.96.047209} would result in a $\sim 250$~mT splitting of the local field at the muon site, much larger than the observed width of 50~mT; however this proposed modulation was not identified in a recent magnetic pair distribution function study \cite{Frandsen:vk5003}.

In the high-temperature cubic phases, modelled using the same spin texture but constraining the lattice angles to 90$^{\circ}$, the situation is reversed:
the lowest energy sites are the ones labeled 2 and 3  in Fig.~\ref{fig:sites} (the green spheres, sites 1 are 32 meV higher in energy, see Table~\ref{tab:muons} and SI~\cite{Note1}).
These sites have almost zero hyperfine contact term (estimated from DFT) and different
$\mathbf{B}_{\mathrm{dip}}$ contributions. Notice that the average local field experienced by a muon hopping between sites 1, or 3, or among sites 2 would vanish. The average field would vanish \textit{a fortiori} when hopping among all three kinds of sites. This clearly does not happen in the  rhombohedral structure up to $T_N$, since the internal field is detected up to the transition. 
The energy difference between geometrically equivalent sites is a direct consequence of the magnetic order that breaks the cubic symmetry and induces 
a substantial electron-density redistribution \cite{PhysRevLett.82.430} observed experimentally \cite{PhysRevB.67.184420,Vidal2002} even in the high-temperature cubic phase \footnote{In a spin unpolarized simulation the energy difference vanishes.}.
Therefore, contrary to naive expectations, the muon local energy landscape is ultimately dictated by the magnetic exchange interactions, despite the small energy associated with magnetostriction \cite{Lee2016}. The effect is highly non-trivial since it manifests itself in a combination of magnetostrictive lattice distortion and charge order. The latter lifts the degeneracy of the eight equivalent muon sites in the cubic crystal, and even makes some of them (the green ones in Fig.~\ref{fig:sites}) unstable in the rhombohedral structure.
This assignment reconciles the observed \msr signal with the general expectation that positive muons occupy positions close to electronegative atoms.

\begin{figure}
    \includegraphics[]{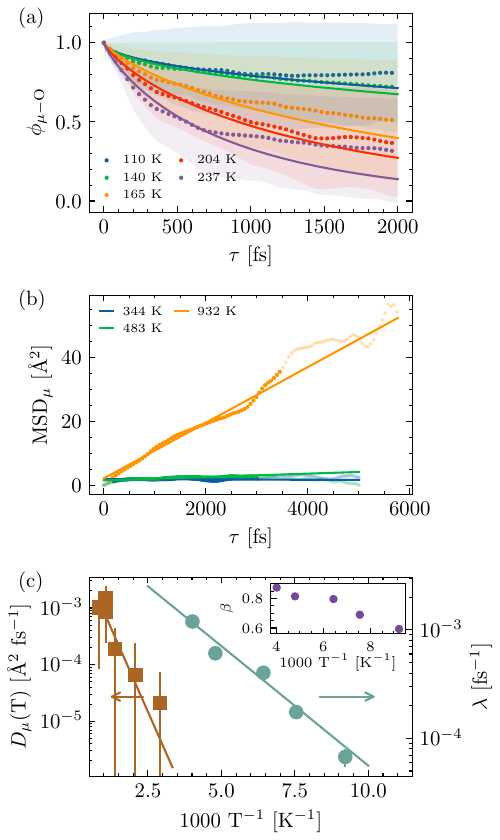}
    \caption{Results from MD simulations. (a)~Angular autocorrelation function, Eq.~\eqref{eq:phi}, for a muon bound to a oxygen, with solid fit curves (see text). Shaded areas display the uncertainty. (b)~The MSD of the muon. Solid lines are linear fit to the data highlighted by opaque points. (c)~Muon diffusion coefficient (brown squares) and decay rate of the angular autocorrelation function (turquoise circles) as a function of the inverse temperature, together with fits to Arrhenius equations (see main text). Finally, the inset shows the values obtained for the exponent of the stretched exponential fits shown in (a). \label{fig:md}}
\end{figure}

Having clarified the description of the ZF results in the magnetically ordered phase, we focus now on the temperature dependence of the ZF depolarization rate and on the Knight shift in the paramagnetic phase. As already mentioned, the peculiar time-dependent behavior of the Knight shift 
\cite{hayanoLongitudinalSpinRelaxation1978, Ishida1984} and the slow temperature variation of the ZF relaxation rates observed in the range $T_{N} < T < 300$~K can be interpreted using a diffusion model that assumes single hops to a second ``site'' with a different Knight shift \cite{uemuraSpinRelaxationPositive1979}. In addition, a second jump in the ZF depolarization rate at 540~K was reported 
%was however reported above 540~K by Lidstr\"om and Hartmann 
\cite{lidstromParamagneticFluctuationsMnO2000a}.
Both effects can be understood in the light of multiple interstitial sites present in the cubic phase around the oxygen atoms, all separated by small energy barriers (see SI \cite{Note1}).
A thermally activated delocalization of the muon on the energy minima around the oxygen forms in the cubic phase: This new state is expected to show zero dipolar contribution to the Knight shift for symmetry reasons. At even higher temperatures the muon will diffuse incoherently throughout the crystal and possibly reach Mn vacancies where a different Knight shift is probed. We therefore expect two distinct dynamic processes, a locally confined dynamics around a single O atom and a second regime  of classical hopping among sites coordinated to different O atoms, thereby explaining
both experimental observations.
A quantitative estimation of the consequent time dependent Knight shift is hardly possible owing to the large uncertainty in the impurity concentration \cite{KOFSTAD1983879,10.1063/1.1712241} and the parameters of the diffusion process reported below.

An accurate description of the muon states as a function of temperature, which involves also the analysis of the structural transition, is beyond the scope of the present work, since it requires detailed, computationally very intensive estimations of the vibrational contributions to the free energy (from both the lattice and the muon) as a function of temperature.
However, qualitative support for the existence of two states may be obtained with molecular dynamics (MD) simulations, using machine learned force fields trained from \textit{ab initio} MD in the cubic symmetry. 
Classical MD is inappropriate at low temperatures owing to the small mass of the muon (about 1/9 of the proton), but the quantum contribution to the dynamics becomes progressively less relevant at higher temperature.

A 64-atom supercell plus the muon is initialized in the lowest energy configuration, assigned random velocities according to the Maxwell–Boltzmann distribution, thermalized using a Nos\'e-Hoover thermostat for 50~fs, 
and eventually evolved with microcanonical dynamics 
to rule out any possible influence of the thermostat on the dynamics.
From the MD trajectories we first compute the angular autocorrelation function $\phi_{\mu - \mathrm{O}}(\tau)=\langle\hat{\mathbf{r}}(t) \cdot \hat{\mathbf{r}}(t+\tau)\rangle$
for the muon bound to an oxygen using
\begin{equation}
    \label{eq:phi}
    \phi_{\mu - \mathrm{O}}(\tau)= \frac{1}{N_{\text {s }}-n_{\tau}}  \sum_{j=1}^{N_{\text {s}}-n_{\tau}} \mathbf{\hat r}_{\mu-O}\left(t_{j}+\tau \right) \cdot \mathbf{\hat r}_{\mu-O}\left(t_{j}\right) \,,
\end{equation}
where $\mathbf{\hat r}_{\mu - O}$ is the unit vector joining the muon
and the oxygen atom it binds to, $N_{\text{s}}$ is the total amount of
molecular dynamics steps and $n_{\tau} = \tau / \Delta t$ where $\Delta t$ is a time step of the simulation.
The maximum $n_\tau$ is set to $N_{\text{s}}/2$ in Eq.~\eqref{eq:phi} in order to accumulate enough statistics and thus reduce the uncertainty \cite{PhysRevA.36.958} of the autocorrelation function.

Representative curves are shown in Fig.~\ref{fig:md}(a), while results presented in panel (c) are obtained with additional averages obtained over 4 realizations.
As it can be appreciated in Fig.~\ref{fig:md}(a), the angular autocorrelation
function decreases rapidly with temperature and its trend is
captured with fits to a stretched exponential $\phi_{\mu - \mathrm{O}}(\tau) = \exp(- (\lambda t)^{\beta})$ which can reproduce
the long time tail (for $t\to0$, $\mu$--O vibrations induce fast oscillations \cite{mattoniMethylammoniumRotationalDynamics2015}). The values of $\lambda$ and $\beta$ are shown in Fig.~\ref{fig:md}(c) as a function of the inverse temperature.
The muon mean square displacement, MSD$_\mu$, is instead obtained from \emph{ab initio} MD and is estimated as
\begin{equation}
\operatorname{MSD}_{\mathrm{\mu}}(\tau)= \frac{1}{N_{\text {s }}-n_{\tau}}  \sum_{j=1}^{N_{\text {s }}-n_{\tau}}\left|\mathbf{r}_{\mu}\left(t_{j}+\tau\right)-\mathbf{r}_{\mu}\left(t_{j}\right)\right|^{2} \,,\label{eq:msd}
\end{equation}
and the diffusion coefficient, defined by
%\begin{equation}
$D_{\mu}=\lim _{\tau \rightarrow \infty} \operatorname{MSD}_{\mu}(\tau)/(6 \tau)    $,
%\end{equation}
is obtained from the MSD as the slope of the linear fits in the region highlighted in Fig.~\ref{fig:md}(b), 
obtained following Ref.~\onlinecite{PhysRev.182.280}, to ensure both linearity of the MSD at large $\tau$ \cite{He2018} and exclusion of ballistic regime at small $\tau$.
Both 
$\lambda$ 
%$\phi_{\mu - \mathrm{O}}$ 
and $D_{\mu}$ follow an activated behavior
\begin{equation}
    D_\mu(T) = D_{0} e ^{-E_{A}^{D}/k_{B}T}, \quad 
    \lambda(T) = \lambda_{0} e ^{-E_{A}^{\phi}/k_{B}T},
\end{equation}
as shown in Fig.~\ref{fig:md}(c). 
The activation energy obtained for the two processes is $E_{A}^{\phi}\sim 44(1)$~meV and $E_{A}^{D}\sim 0.23(16)$~eV and the infinite-temperature limits are $\lambda_0=0.01(1)$~fs$^{-1}$ and $D_{0}=0.01(3)$~\AA$^2$fs$^{-1}$. The results for the former process can be roughly compared with the experimental estimates for the activation energy $E_A^{\text{exp}}=61$~meV and pre-exponential time $1/\tau_{0} ^{\text{exp}}=0.169$~fs$^{-1}$ of an activated hopping process \footnote{The experimental estimate includes a correction due to the electronic correlation time, see cited reference for details.}\cite{uemuraSpinRelaxationPositive1979}.
Most importantly, the MD simulations show the presence of two diffusion mechanisms, one localized around oxygen atoms leading to the effective new site resulting from the motion average and experimentally observed above the magnetic transition and a second, conventional, diffusion process among oxygen atoms that justifies the second transition observed at high temperatures.
We stress that the quantitative predictions should be taken \emph{cum grano salis}, since the quantum contributions to the muon motion are expected to be far from negligible in the temperature interval of rotational diffusion.

In conclusion, we have clarified the interpretation of the \msr{} signal of MnO---a puzzle that remained unsolved for over 40 years.
The solution is naturally obtained from the accurate description of the magnetostriction effect that plays a fundamental role not only in the structural phase transition but also in the stabilization of different interstitial sites in the two phases of MnO. This leads to the observation of a single precession frequency below $T_N$ and explains the unusual behaviours observed at higher temperatures. Indeed the particular network of muon sites stabilized in the cubic phase gives rise to two diffusion regimes, qualitatively captured by our MD simulations, that explain the low temperature time dependence of the Knight shift and the second diffusion process observed experimentally above 500~K.

Our results highlight the fact that relatively small energy differences can play a fundamental role in the determination of the muon-sample interaction.
This conclusion is applicable to the magnetic state of other transition-metal oxides whose \msr interpretation is still lacking \cite{Nishiyama2001, Nishiyama1997} and provides an important ingredient to be considered in the analysis of experimental results of magnetic oxides that have recently attracted scientific attention owing to muon-induced effects \cite{PhysRevLett.126.037202,PhysRevX.10.011036}.
%An accurate description of the electronic interactions in the system turns out to be fundamental not only to account for the complicated Mott-Hubbard ground state but also to correctly identify the energy scales among the various geometric configurations that in turn define the interstitial sites occupied by the muon and therefore the experimental signal. 

PB, IO and RDR acknowledge financial support from PNRR MUR project PE0000023-NQSTI. IT, LM, GP, and NM acknowledge financial support from NCCR MARVEL, a National Centre of Competence in Research, funded by the Swiss National Science Foundation (grant number 205602). 
SJB was funded by UK Research and Innovation (UKRI) under the UK government’s Horizon Europe funding guarantee [grant number EP/X025861/1].
Computer time was provided by the Swiss National Supercomputing Centre (CSCS) under projects No.~s1073 and s1192 and by the STFC Scientific Computing Department's SCARF cluster.
The authors acknowledge Andrew Goodwin and Matthias Gutmann for sending their sample, sharing experimental detail and useful discussions, and Giuliana Materzanini for useful discussion on the MD analysis.

\bibliographystyle{apsrev4-1}
\bibliography{main}

%merlin.mbs apsrev4-1.bst 2010-07-25 4.21a (PWD, AO, DPC) hacked
%Control: key (0)
%Control: author (72) initials jnrlst
%Control: editor formatted (1) identically to author
%Control: production of article title (-1) disabled
%Control: page (0) single
%Control: year (1) truncated
%Control: production of eprint (0) enabled
\begin{thebibliography}{92}%
\makeatletter
\providecommand \@ifxundefined [1]{%
 \@ifx{#1\undefined}
}%
\providecommand \@ifnum [1]{%
 \ifnum #1\expandafter \@firstoftwo
 \else \expandafter \@secondoftwo
 \fi
}%
\providecommand \@ifx [1]{%
 \ifx #1\expandafter \@firstoftwo
 \else \expandafter \@secondoftwo
 \fi
}%
\providecommand \natexlab [1]{#1}%
\providecommand \enquote  [1]{``#1''}%
\providecommand \bibnamefont  [1]{#1}%
\providecommand \bibfnamefont [1]{#1}%
\providecommand \citenamefont [1]{#1}%
\providecommand \href@noop [0]{\@secondoftwo}%
\providecommand \href [0]{\begingroup \@sanitize@url \@href}%
\providecommand \@href[1]{\@@startlink{#1}\@@href}%
\providecommand \@@href[1]{\endgroup#1\@@endlink}%
\providecommand \@sanitize@url [0]{\catcode `\\12\catcode `\$12\catcode
  `\&12\catcode `\#12\catcode `\^12\catcode `\_12\catcode `\%12\relax}%
\providecommand \@@startlink[1]{}%
\providecommand \@@endlink[0]{}%
\providecommand \url  [0]{\begingroup\@sanitize@url \@url }%
\providecommand \@url [1]{\endgroup\@href {#1}{\urlprefix }}%
\providecommand \urlprefix  [0]{URL }%
\providecommand \Eprint [0]{\href }%
\providecommand \doibase [0]{http://dx.doi.org/}%
\providecommand \selectlanguage [0]{\@gobble}%
\providecommand \bibinfo  [0]{\@secondoftwo}%
\providecommand \bibfield  [0]{\@secondoftwo}%
\providecommand \translation [1]{[#1]}%
\providecommand \BibitemOpen [0]{}%
\providecommand \bibitemStop [0]{}%
\providecommand \bibitemNoStop [0]{.\EOS\space}%
\providecommand \EOS [0]{\spacefactor3000\relax}%
\providecommand \BibitemShut  [1]{\csname bibitem#1\endcsname}%
\let\auto@bib@innerbib\@empty
%</preamble>
\bibitem [{\citenamefont {Blundell}\ \emph {et~al.}(2022)\citenamefont
  {Blundell}, \citenamefont {Renzi}, \citenamefont {Lancaster},\ and\
  \citenamefont {Pratt}}]{blundell2022}%
  \BibitemOpen
  \bibinfo {editor} {\bibfnamefont {S.~J.}\ \bibnamefont {Blundell}}, \bibinfo
  {editor} {\bibfnamefont {R.~D.}\ \bibnamefont {Renzi}}, \bibinfo {editor}
  {\bibfnamefont {T.}~\bibnamefont {Lancaster}}, \ and\ \bibinfo {editor}
  {\bibfnamefont {F.~L.}\ \bibnamefont {Pratt}},\ eds.,\ \href@noop {} {\emph
  {\bibinfo {title} {Muon Spectroscopy - An Introduction}}}\ (\bibinfo
  {publisher} {Oxford University Press},\ \bibinfo {address} {Oxford},\
  \bibinfo {year} {2022})\BibitemShut {NoStop}%
\bibitem [{\citenamefont {Roth}(1958)}]{rothMagneticStructuresMnO1958}%
  \BibitemOpen
  \bibfield  {author} {\bibinfo {author} {\bibfnamefont {W.~L.}\ \bibnamefont
  {Roth}},\ }\href {\doibase 10.1103/PhysRev.110.1333} {\bibfield  {journal}
  {\bibinfo  {journal} {Physical Review}\ }\textbf {\bibinfo {volume} {110}},\
  \bibinfo {pages} {1333} (\bibinfo {year} {1958})}\BibitemShut {NoStop}%
\bibitem [{\citenamefont {Shaked}\ \emph {et~al.}(1988)\citenamefont {Shaked},
  \citenamefont {Faber},\ and\ \citenamefont {Hitterman}}]{PhysRevB.38.11901}%
  \BibitemOpen
  \bibfield  {author} {\bibinfo {author} {\bibfnamefont {H.}~\bibnamefont
  {Shaked}}, \bibinfo {author} {\bibfnamefont {J.}~\bibnamefont {Faber}}, \
  and\ \bibinfo {author} {\bibfnamefont {R.~L.}\ \bibnamefont {Hitterman}},\
  }\href {\doibase 10.1103/PhysRevB.38.11901} {\bibfield  {journal} {\bibinfo
  {journal} {Phys. Rev. B}\ }\textbf {\bibinfo {volume} {38}},\ \bibinfo
  {pages} {11901} (\bibinfo {year} {1988})}\BibitemShut {NoStop}%
\bibitem [{\citenamefont {Cheetham}\ and\ \citenamefont
  {Hope}(1983)}]{PhysRevB.27.6964}%
  \BibitemOpen
  \bibfield  {author} {\bibinfo {author} {\bibfnamefont {A.~K.}\ \bibnamefont
  {Cheetham}}\ and\ \bibinfo {author} {\bibfnamefont {D.~A.~O.}\ \bibnamefont
  {Hope}},\ }\href {\doibase 10.1103/PhysRevB.27.6964} {\bibfield  {journal}
  {\bibinfo  {journal} {Phys. Rev. B}\ }\textbf {\bibinfo {volume} {27}},\
  \bibinfo {pages} {6964} (\bibinfo {year} {1983})}\BibitemShut {NoStop}%
\bibitem [{\citenamefont {Goodwin}\ \emph {et~al.}(2006)\citenamefont
  {Goodwin}, \citenamefont {Tucker}, \citenamefont {Dove},\ and\ \citenamefont
  {Keen}}]{PhysRevLett.96.047209}%
  \BibitemOpen
  \bibfield  {author} {\bibinfo {author} {\bibfnamefont {A.~L.}\ \bibnamefont
  {Goodwin}}, \bibinfo {author} {\bibfnamefont {M.~G.}\ \bibnamefont {Tucker}},
  \bibinfo {author} {\bibfnamefont {M.~T.}\ \bibnamefont {Dove}}, \ and\
  \bibinfo {author} {\bibfnamefont {D.~A.}\ \bibnamefont {Keen}},\ }\href
  {\doibase 10.1103/PhysRevLett.96.047209} {\bibfield  {journal} {\bibinfo
  {journal} {Phys. Rev. Lett.}\ }\textbf {\bibinfo {volume} {96}},\ \bibinfo
  {pages} {047209} (\bibinfo {year} {2006})}\BibitemShut {NoStop}%
\bibitem [{\citenamefont {Gazzara}\ and\ \citenamefont
  {Middleton}(1965)}]{gazzara_middleton_1965}%
  \BibitemOpen
  \bibfield  {author} {\bibinfo {author} {\bibfnamefont {C.~P.}\ \bibnamefont
  {Gazzara}}\ and\ \bibinfo {author} {\bibfnamefont {R.~M.}\ \bibnamefont
  {Middleton}},\ }\href {\doibase 10.1154/S0376030800003499} {\bibfield
  {journal} {\bibinfo  {journal} {Advances in X-Ray Analysis}\ }\textbf
  {\bibinfo {volume} {9}},\ \bibinfo {pages} {152–158} (\bibinfo {year}
  {1965})}\BibitemShut {NoStop}%
\bibitem [{\citenamefont {Srinivasan}\ and\ \citenamefont
  {Seehra}(1983)}]{PhysRevB.28.6542}%
  \BibitemOpen
  \bibfield  {author} {\bibinfo {author} {\bibfnamefont {G.}~\bibnamefont
  {Srinivasan}}\ and\ \bibinfo {author} {\bibfnamefont {M.~S.}\ \bibnamefont
  {Seehra}},\ }\href {\doibase 10.1103/PhysRevB.28.6542} {\bibfield  {journal}
  {\bibinfo  {journal} {Phys. Rev. B}\ }\textbf {\bibinfo {volume} {28}},\
  \bibinfo {pages} {6542} (\bibinfo {year} {1983})}\BibitemShut {NoStop}%
\bibitem [{\citenamefont {Yamamoto}\ and\ \citenamefont
  {Nagamiya}(1972)}]{yamamotoSpinArrangementsMagnetic1972}%
  \BibitemOpen
  \bibfield  {author} {\bibinfo {author} {\bibfnamefont {Y.}~\bibnamefont
  {Yamamoto}}\ and\ \bibinfo {author} {\bibfnamefont {T.}~\bibnamefont
  {Nagamiya}},\ }\href {\doibase 10.1143/JPSJ.32.1248} {\bibfield  {journal}
  {\bibinfo  {journal} {Journal of the Physical Society of Japan}\ }\textbf
  {\bibinfo {volume} {32}},\ \bibinfo {pages} {1248} (\bibinfo {year}
  {1972})}\BibitemShut {NoStop}%
\bibitem [{\citenamefont {Blech}\ and\ \citenamefont
  {Averbach}(1966)}]{PhysRev.142.287}%
  \BibitemOpen
  \bibfield  {author} {\bibinfo {author} {\bibfnamefont {I.~A.}\ \bibnamefont
  {Blech}}\ and\ \bibinfo {author} {\bibfnamefont {B.~L.}\ \bibnamefont
  {Averbach}},\ }\href {\doibase 10.1103/PhysRev.142.287} {\bibfield  {journal}
  {\bibinfo  {journal} {Phys. Rev.}\ }\textbf {\bibinfo {volume} {142}},\
  \bibinfo {pages} {287} (\bibinfo {year} {1966})}\BibitemShut {NoStop}%
\bibitem [{\citenamefont {Uchida}\ \emph {et~al.}(1960)\citenamefont {Uchida},
  \citenamefont {Kondoh}, \citenamefont {Nakazumi},\ and\ \citenamefont
  {Nagamiya}}]{10.1143/JPSJ.15.466}%
  \BibitemOpen
  \bibfield  {author} {\bibinfo {author} {\bibfnamefont {E.}~\bibnamefont
  {Uchida}}, \bibinfo {author} {\bibfnamefont {H.}~\bibnamefont {Kondoh}},
  \bibinfo {author} {\bibfnamefont {Y.}~\bibnamefont {Nakazumi}}, \ and\
  \bibinfo {author} {\bibfnamefont {T.}~\bibnamefont {Nagamiya}},\ }\href
  {\doibase 10.1143/JPSJ.15.466} {\bibfield  {journal} {\bibinfo  {journal}
  {Journal of the Physical Society of Japan}\ }\textbf {\bibinfo {volume}
  {15}},\ \bibinfo {pages} {466} (\bibinfo {year} {1960})}\BibitemShut
  {NoStop}%
\bibitem [{\citenamefont {Morosin}(1970)}]{PhysRevB.1.236}%
  \BibitemOpen
  \bibfield  {author} {\bibinfo {author} {\bibfnamefont {B.}~\bibnamefont
  {Morosin}},\ }\href {\doibase 10.1103/PhysRevB.1.236} {\bibfield  {journal}
  {\bibinfo  {journal} {Phys. Rev. B}\ }\textbf {\bibinfo {volume} {1}},\
  \bibinfo {pages} {236} (\bibinfo {year} {1970})}\BibitemShut {NoStop}%
\bibitem [{\citenamefont {Balagurov}\ \emph {et~al.}(2016)\citenamefont
  {Balagurov}, \citenamefont {Bobrikov}, \citenamefont {Sumnikov},
  \citenamefont {Yushankhai},\ and\ \citenamefont
  {{Mironova-Ulmane}}}]{balagurovMagnetostructuralPhaseTransitions2016}%
  \BibitemOpen
  \bibfield  {author} {\bibinfo {author} {\bibfnamefont {A.~M.}\ \bibnamefont
  {Balagurov}}, \bibinfo {author} {\bibfnamefont {I.~A.}\ \bibnamefont
  {Bobrikov}}, \bibinfo {author} {\bibfnamefont {S.~V.}\ \bibnamefont
  {Sumnikov}}, \bibinfo {author} {\bibfnamefont {V.~Y.}\ \bibnamefont
  {Yushankhai}}, \ and\ \bibinfo {author} {\bibfnamefont {N.}~\bibnamefont
  {{Mironova-Ulmane}}},\ }\href {\doibase 10.1134/S0021364016140071} {\bibfield
   {journal} {\bibinfo  {journal} {JETP Letters}\ }\textbf {\bibinfo {volume}
  {104}},\ \bibinfo {pages} {88} (\bibinfo {year} {2016})}\BibitemShut
  {NoStop}%
\bibitem [{\citenamefont {Uemura}\ \emph
  {et~al.}(1981{\natexlab{a}})\citenamefont {Uemura}, \citenamefont {Nishida},
  \citenamefont {Imazato}, \citenamefont {Hayano}, \citenamefont {Takigawa},\
  and\ \citenamefont {Yamazaki}}]{Uemura1981b}%
  \BibitemOpen
  \bibfield  {author} {\bibinfo {author} {\bibfnamefont {Y.~J.}\ \bibnamefont
  {Uemura}}, \bibinfo {author} {\bibfnamefont {N.}~\bibnamefont {Nishida}},
  \bibinfo {author} {\bibfnamefont {J.}~\bibnamefont {Imazato}}, \bibinfo
  {author} {\bibfnamefont {R.~S.}\ \bibnamefont {Hayano}}, \bibinfo {author}
  {\bibfnamefont {M.}~\bibnamefont {Takigawa}}, \ and\ \bibinfo {author}
  {\bibfnamefont {T.}~\bibnamefont {Yamazaki}},\ }\href {\doibase
  10.1007/BF01037559} {\bibfield  {journal} {\bibinfo  {journal} {Hyperfine
  Interact}\ }\textbf {\bibinfo {volume} {8}},\ \bibinfo {pages} {771}
  (\bibinfo {year} {1981}{\natexlab{a}})}\BibitemShut {NoStop}%
\bibitem [{\citenamefont {Uemura}\ \emph {et~al.}(1984)\citenamefont {Uemura},
  \citenamefont {Yamazaki}, \citenamefont {Kitaoka}, \citenamefont {Takigawa},\
  and\ \citenamefont {Yasuoka}}]{Uemura1984}%
  \BibitemOpen
  \bibfield  {author} {\bibinfo {author} {\bibfnamefont {Y.~J.}\ \bibnamefont
  {Uemura}}, \bibinfo {author} {\bibfnamefont {T.}~\bibnamefont {Yamazaki}},
  \bibinfo {author} {\bibfnamefont {Y.}~\bibnamefont {Kitaoka}}, \bibinfo
  {author} {\bibfnamefont {M.}~\bibnamefont {Takigawa}}, \ and\ \bibinfo
  {author} {\bibfnamefont {H.}~\bibnamefont {Yasuoka}},\ }\href {\doibase
  10.1007/BF02065922} {\bibfield  {journal} {\bibinfo  {journal} {Hyperfine
  Interact}\ }\textbf {\bibinfo {volume} {17}},\ \bibinfo {pages} {339}
  (\bibinfo {year} {1984})}\BibitemShut {NoStop}%
\bibitem [{\citenamefont {Uemura}\ \emph {et~al.}(1979)\citenamefont {Uemura},
  \citenamefont {Hayano}, \citenamefont {Imazato}, \citenamefont {Nishida},
  \citenamefont {Nagamine}, \citenamefont {Yamazaki},\ and\ \citenamefont
  {Yasuoka}}]{uemuraSpinRelaxationPositive1979}%
  \BibitemOpen
  \bibfield  {author} {\bibinfo {author} {\bibfnamefont {Y.~J.}\ \bibnamefont
  {Uemura}}, \bibinfo {author} {\bibfnamefont {R.~S.}\ \bibnamefont {Hayano}},
  \bibinfo {author} {\bibfnamefont {J.}~\bibnamefont {Imazato}}, \bibinfo
  {author} {\bibfnamefont {N.}~\bibnamefont {Nishida}}, \bibinfo {author}
  {\bibfnamefont {K.}~\bibnamefont {Nagamine}}, \bibinfo {author}
  {\bibfnamefont {T.}~\bibnamefont {Yamazaki}}, \ and\ \bibinfo {author}
  {\bibfnamefont {H.}~\bibnamefont {Yasuoka}},\ }\href {\doibase
  10.1007/BF01028780} {\bibfield  {journal} {\bibinfo  {journal} {Hyperfine
  Interact}\ }\textbf {\bibinfo {volume} {6}},\ \bibinfo {pages} {127}
  (\bibinfo {year} {1979})}\BibitemShut {NoStop}%
\bibitem [{\citenamefont {Uemura}\ \emph
  {et~al.}(1981{\natexlab{b}})\citenamefont {Uemura}, \citenamefont {Imazato},
  \citenamefont {Nishida}, \citenamefont {Hayano}, \citenamefont {Takigawa},\
  and\ \citenamefont {Yamazaki}}]{Uemura1981a}%
  \BibitemOpen
  \bibfield  {author} {\bibinfo {author} {\bibfnamefont {Y.~J.}\ \bibnamefont
  {Uemura}}, \bibinfo {author} {\bibfnamefont {J.}~\bibnamefont {Imazato}},
  \bibinfo {author} {\bibfnamefont {N.}~\bibnamefont {Nishida}}, \bibinfo
  {author} {\bibfnamefont {R.~S.}\ \bibnamefont {Hayano}}, \bibinfo {author}
  {\bibfnamefont {M.}~\bibnamefont {Takigawa}}, \ and\ \bibinfo {author}
  {\bibfnamefont {T.}~\bibnamefont {Yamazaki}},\ }\href {\doibase
  10.1007/BF01037552} {\bibfield  {journal} {\bibinfo  {journal} {Hyperfine
  Interact}\ }\textbf {\bibinfo {volume} {8}},\ \bibinfo {pages} {725}
  (\bibinfo {year} {1981}{\natexlab{b}})}\BibitemShut {NoStop}%
\bibitem [{\citenamefont {Ishida}\ \emph {et~al.}(1984)\citenamefont {Ishida},
  \citenamefont {Matsuzaki}, \citenamefont {Nishiyama},\ and\ \citenamefont
  {Nagamine}}]{Ishida1984}%
  \BibitemOpen
  \bibfield  {author} {\bibinfo {author} {\bibfnamefont {K.}~\bibnamefont
  {Ishida}}, \bibinfo {author} {\bibfnamefont {T.}~\bibnamefont {Matsuzaki}},
  \bibinfo {author} {\bibfnamefont {K.}~\bibnamefont {Nishiyama}}, \ and\
  \bibinfo {author} {\bibfnamefont {K.}~\bibnamefont {Nagamine}},\ }\href
  {\doibase 10.1007/BF02066138} {\bibfield  {journal} {\bibinfo  {journal}
  {Hyperfine Interact}\ }\textbf {\bibinfo {volume} {19}},\ \bibinfo {pages}
  {927} (\bibinfo {year} {1984})}\BibitemShut {NoStop}%
\bibitem [{\citenamefont {Aschauer}\ \emph {et~al.}(2015)\citenamefont
  {Aschauer}, \citenamefont {Vonr{\"u}ti},\ and\ \citenamefont
  {Spaldin}}]{aschauerEffectEpitaxialStrain2015}%
  \BibitemOpen
  \bibfield  {author} {\bibinfo {author} {\bibfnamefont {U.}~\bibnamefont
  {Aschauer}}, \bibinfo {author} {\bibfnamefont {N.}~\bibnamefont
  {Vonr{\"u}ti}}, \ and\ \bibinfo {author} {\bibfnamefont {N.~A.}\ \bibnamefont
  {Spaldin}},\ }\href {\doibase 10.1103/PhysRevB.92.054103} {\bibfield
  {journal} {\bibinfo  {journal} {Physical Review B}\ }\textbf {\bibinfo
  {volume} {92}},\ \bibinfo {pages} {054103} (\bibinfo {year}
  {2015})}\BibitemShut {NoStop}%
\bibitem [{\citenamefont {Logsdail}\ \emph {et~al.}(2019)\citenamefont
  {Logsdail}, \citenamefont {Downing}, \citenamefont {Keal}, \citenamefont
  {Sherwood}, \citenamefont {Sokol},\ and\ \citenamefont
  {Catlow}}]{logsdailHybridDFTModelingLattice2019}%
  \BibitemOpen
  \bibfield  {author} {\bibinfo {author} {\bibfnamefont {A.~J.}\ \bibnamefont
  {Logsdail}}, \bibinfo {author} {\bibfnamefont {C.~A.}\ \bibnamefont
  {Downing}}, \bibinfo {author} {\bibfnamefont {T.~W.}\ \bibnamefont {Keal}},
  \bibinfo {author} {\bibfnamefont {P.}~\bibnamefont {Sherwood}}, \bibinfo
  {author} {\bibfnamefont {A.~A.}\ \bibnamefont {Sokol}}, \ and\ \bibinfo
  {author} {\bibfnamefont {C.~R.~A.}\ \bibnamefont {Catlow}},\ }\href {\doibase
  10.1021/acs.jpcc.8b07846} {\bibfield  {journal} {\bibinfo  {journal} {The
  Journal of Physical Chemistry C}\ }\textbf {\bibinfo {volume} {123}},\
  \bibinfo {pages} {8133} (\bibinfo {year} {2019})}\BibitemShut {NoStop}%
\bibitem [{\citenamefont {Hayano}\ \emph {et~al.}(1978)\citenamefont {Hayano},
  \citenamefont {Uemura}, \citenamefont {Imazato}, \citenamefont {Nishida},
  \citenamefont {Nagamine}, \citenamefont {Yamazaki},\ and\ \citenamefont
  {Yasuoka}}]{hayanoLongitudinalSpinRelaxation1978}%
  \BibitemOpen
  \bibfield  {author} {\bibinfo {author} {\bibfnamefont {R.~S.}\ \bibnamefont
  {Hayano}}, \bibinfo {author} {\bibfnamefont {Y.~J.}\ \bibnamefont {Uemura}},
  \bibinfo {author} {\bibfnamefont {J.}~\bibnamefont {Imazato}}, \bibinfo
  {author} {\bibfnamefont {N.}~\bibnamefont {Nishida}}, \bibinfo {author}
  {\bibfnamefont {K.}~\bibnamefont {Nagamine}}, \bibinfo {author}
  {\bibfnamefont {T.}~\bibnamefont {Yamazaki}}, \ and\ \bibinfo {author}
  {\bibfnamefont {H.}~\bibnamefont {Yasuoka}},\ }\href {\doibase
  10.1103/PhysRevLett.41.421} {\bibfield  {journal} {\bibinfo  {journal}
  {Physical Review Letters}\ }\textbf {\bibinfo {volume} {41}},\ \bibinfo
  {pages} {421} (\bibinfo {year} {1978})}\BibitemShut {NoStop}%
\bibitem [{\citenamefont {Lidstr{\"o}m}\ and\ \citenamefont
  {Hartmann}(2000)}]{lidstromParamagneticFluctuationsMnO2000a}%
  \BibitemOpen
  \bibfield  {author} {\bibinfo {author} {\bibfnamefont {E.}~\bibnamefont
  {Lidstr{\"o}m}}\ and\ \bibinfo {author} {\bibfnamefont {O.}~\bibnamefont
  {Hartmann}},\ }\href {\doibase 10.1088/0953-8984/12/23/306} {\bibfield
  {journal} {\bibinfo  {journal} {J. Phys.: Condens. Matter}\ }\textbf
  {\bibinfo {volume} {12}},\ \bibinfo {pages} {4969} (\bibinfo {year}
  {2000})}\BibitemShut {NoStop}%
\bibitem [{\citenamefont {Hermsmeier}\ \emph {et~al.}(1989)\citenamefont
  {Hermsmeier}, \citenamefont {Osterwalder}, \citenamefont {Friedman},\ and\
  \citenamefont {Fadley}}]{PhysRevLett.62.478}%
  \BibitemOpen
  \bibfield  {author} {\bibinfo {author} {\bibfnamefont {B.}~\bibnamefont
  {Hermsmeier}}, \bibinfo {author} {\bibfnamefont {J.}~\bibnamefont
  {Osterwalder}}, \bibinfo {author} {\bibfnamefont {D.~J.}\ \bibnamefont
  {Friedman}}, \ and\ \bibinfo {author} {\bibfnamefont {C.~S.}\ \bibnamefont
  {Fadley}},\ }\href {\doibase 10.1103/PhysRevLett.62.478} {\bibfield
  {journal} {\bibinfo  {journal} {Phys. Rev. Lett.}\ }\textbf {\bibinfo
  {volume} {62}},\ \bibinfo {pages} {478} (\bibinfo {year} {1989})}\BibitemShut
  {NoStop}%
\bibitem [{\citenamefont {Hermsmeier}\ \emph {et~al.}(1990)\citenamefont
  {Hermsmeier}, \citenamefont {Osterwalder}, \citenamefont {Friedman},
  \citenamefont {Sinkovic}, \citenamefont {Tran},\ and\ \citenamefont
  {Fadley}}]{PhysRevB.42.11895}%
  \BibitemOpen
  \bibfield  {author} {\bibinfo {author} {\bibfnamefont {B.}~\bibnamefont
  {Hermsmeier}}, \bibinfo {author} {\bibfnamefont {J.}~\bibnamefont
  {Osterwalder}}, \bibinfo {author} {\bibfnamefont {D.~J.}\ \bibnamefont
  {Friedman}}, \bibinfo {author} {\bibfnamefont {B.}~\bibnamefont {Sinkovic}},
  \bibinfo {author} {\bibfnamefont {T.}~\bibnamefont {Tran}}, \ and\ \bibinfo
  {author} {\bibfnamefont {C.~S.}\ \bibnamefont {Fadley}},\ }\href {\doibase
  10.1103/PhysRevB.42.11895} {\bibfield  {journal} {\bibinfo  {journal} {Phys.
  Rev. B}\ }\textbf {\bibinfo {volume} {42}},\ \bibinfo {pages} {11895}
  (\bibinfo {year} {1990})}\BibitemShut {NoStop}%
\bibitem [{\citenamefont {Sun}\ \emph {et~al.}(2017)\citenamefont {Sun},
  \citenamefont {Feng}, \citenamefont {Su}, \citenamefont {Nemkovski},
  \citenamefont {Petracic},\ and\ \citenamefont {Brückel}}]{Sun_2017}%
  \BibitemOpen
  \bibfield  {author} {\bibinfo {author} {\bibfnamefont {X.}~\bibnamefont
  {Sun}}, \bibinfo {author} {\bibfnamefont {E.}~\bibnamefont {Feng}}, \bibinfo
  {author} {\bibfnamefont {Y.}~\bibnamefont {Su}}, \bibinfo {author}
  {\bibfnamefont {K.}~\bibnamefont {Nemkovski}}, \bibinfo {author}
  {\bibfnamefont {O.}~\bibnamefont {Petracic}}, \ and\ \bibinfo {author}
  {\bibfnamefont {T.}~\bibnamefont {Brückel}},\ }\href {\doibase
  10.1088/1742-6596/862/1/012027} {\bibfield  {journal} {\bibinfo  {journal}
  {Journal of Physics: Conference Series}\ }\textbf {\bibinfo {volume} {862}},\
  \bibinfo {pages} {012027} (\bibinfo {year} {2017})}\BibitemShut {NoStop}%
\bibitem [{Note1()}]{Note1}%
  \BibitemOpen
  \bibinfo {note} {See Supplemental Material at [URL will be inserted by
  publisher] for the details of the experimental setup, the computational
  methods, the details of the simulations and the analysis of experimental and
  theoretical data. It also includes Refs.~\cite
  {10.1063/5.0005077,Baroni:2001,BENNETT2019109137,Bonfa2016,Campo:2010,chmiela2023,Cococcioni:2005,DALCORSO2014337,Dudarev:1998,GARRITY2014446,giannozzi2020quantum,Giannozzi_2009,Giannozzi_2017,gipaw-website,Gorni:2018,He2018,Hohenberg:1964,Kohn:1965,Kucukbenli:2014,Kulik:2011,Lowdin:1950,MaterialsCloud,MaterialsCloudArchive2023,Mayer:2002,PhysRev.182.280,PhysRevA.36.958,PhysRevB.1.236,PhysRevB.23.5048,PhysRevB.27.6964,PhysRevB.38.11901,PhysRevB.7.5212,PhysRevB.97.174414,PhysRevLett.100.136406,PhysRevMaterials.3.073804,prandini2018precision,PRATT2000710,Timrov:2018,Timrov:2020b,Timrov:2021,Timrov:2022,Uemura1981a,Wang:2016,Tablero:2008,Amadon:2008,Nawa:2018}.}\BibitemShut
  {Stop}%
\bibitem [{\citenamefont {Hohenberg}\ and\ \citenamefont
  {Kohn}(1964)}]{Hohenberg:1964}%
  \BibitemOpen
  \bibfield  {author} {\bibinfo {author} {\bibfnamefont {P.}~\bibnamefont
  {Hohenberg}}\ and\ \bibinfo {author} {\bibfnamefont {W.}~\bibnamefont
  {Kohn}},\ }\href@noop {} {\bibfield  {journal} {\bibinfo  {journal} {Phys.
  Rev.}\ }\textbf {\bibinfo {volume} {136}},\ \bibinfo {pages} {B864} (\bibinfo
  {year} {1964})}\BibitemShut {NoStop}%
\bibitem [{\citenamefont {Kohn}\ and\ \citenamefont {Sham}(1965)}]{Kohn:1965}%
  \BibitemOpen
  \bibfield  {author} {\bibinfo {author} {\bibfnamefont {W.}~\bibnamefont
  {Kohn}}\ and\ \bibinfo {author} {\bibfnamefont {L.}~\bibnamefont {Sham}},\
  }\href@noop {} {\bibfield  {journal} {\bibinfo  {journal} {Phys. Rev.}\
  }\textbf {\bibinfo {volume} {140}},\ \bibinfo {pages} {A1133} (\bibinfo
  {year} {1965})}\BibitemShut {NoStop}%
\bibitem [{\citenamefont {Anisimov}\ \emph {et~al.}(1991)\citenamefont
  {Anisimov}, \citenamefont {Zaanen},\ and\ \citenamefont
  {Andersen}}]{anisimov:1991}%
  \BibitemOpen
  \bibfield  {author} {\bibinfo {author} {\bibfnamefont {V.~I.}\ \bibnamefont
  {Anisimov}}, \bibinfo {author} {\bibfnamefont {J.}~\bibnamefont {Zaanen}}, \
  and\ \bibinfo {author} {\bibfnamefont {O.~K.}\ \bibnamefont {Andersen}},\
  }\href@noop {} {\bibfield  {journal} {\bibinfo  {journal} {Phys. Rev. B}\
  }\textbf {\bibinfo {volume} {44}},\ \bibinfo {pages} {943} (\bibinfo {year}
  {1991})}\BibitemShut {NoStop}%
\bibitem [{\citenamefont {Dudarev}\ \emph {et~al.}(1998)\citenamefont
  {Dudarev}, \citenamefont {Botton}, \citenamefont {Savrasov}, \citenamefont
  {Humphreys},\ and\ \citenamefont {Sutton}}]{Dudarev:1998}%
  \BibitemOpen
  \bibfield  {author} {\bibinfo {author} {\bibfnamefont {S.~L.}\ \bibnamefont
  {Dudarev}}, \bibinfo {author} {\bibfnamefont {G.~A.}\ \bibnamefont {Botton}},
  \bibinfo {author} {\bibfnamefont {S.~Y.}\ \bibnamefont {Savrasov}}, \bibinfo
  {author} {\bibfnamefont {C.~J.}\ \bibnamefont {Humphreys}}, \ and\ \bibinfo
  {author} {\bibfnamefont {A.~P.}\ \bibnamefont {Sutton}},\ }\href@noop {}
  {\bibfield  {journal} {\bibinfo  {journal} {Physical Review B}\ }\textbf
  {\bibinfo {volume} {57}},\ \bibinfo {pages} {1505} (\bibinfo {year}
  {1998})}\BibitemShut {NoStop}%
\bibitem [{\citenamefont {Campo~Jr}\ and\ \citenamefont
  {Cococcioni}(2010)}]{Campo:2010}%
  \BibitemOpen
  \bibfield  {author} {\bibinfo {author} {\bibfnamefont {V.~L.}\ \bibnamefont
  {Campo~Jr}}\ and\ \bibinfo {author} {\bibfnamefont {M.}~\bibnamefont
  {Cococcioni}},\ }\href@noop {} {\bibfield  {journal} {\bibinfo  {journal} {J.
  Phys.: Condens. Matter}\ }\textbf {\bibinfo {volume} {22}},\ \bibinfo {pages}
  {055602} (\bibinfo {year} {2010})}\BibitemShut {NoStop}%
\bibitem [{\citenamefont {Perdew}\ \emph {et~al.}(2008)\citenamefont {Perdew},
  \citenamefont {Ruzsinszky}, \citenamefont {Csonka}, \citenamefont {Vydrov},
  \citenamefont {Scuseria}, \citenamefont {Constantin}, \citenamefont {Zhou},\
  and\ \citenamefont {Burke}}]{PhysRevLett.100.136406}%
  \BibitemOpen
  \bibfield  {author} {\bibinfo {author} {\bibfnamefont {J.~P.}\ \bibnamefont
  {Perdew}}, \bibinfo {author} {\bibfnamefont {A.}~\bibnamefont {Ruzsinszky}},
  \bibinfo {author} {\bibfnamefont {G.~I.}\ \bibnamefont {Csonka}}, \bibinfo
  {author} {\bibfnamefont {O.~A.}\ \bibnamefont {Vydrov}}, \bibinfo {author}
  {\bibfnamefont {G.~E.}\ \bibnamefont {Scuseria}}, \bibinfo {author}
  {\bibfnamefont {L.~A.}\ \bibnamefont {Constantin}}, \bibinfo {author}
  {\bibfnamefont {X.}~\bibnamefont {Zhou}}, \ and\ \bibinfo {author}
  {\bibfnamefont {K.}~\bibnamefont {Burke}},\ }\href {\doibase
  10.1103/PhysRevLett.100.136406} {\bibfield  {journal} {\bibinfo  {journal}
  {Phys. Rev. Lett.}\ }\textbf {\bibinfo {volume} {100}},\ \bibinfo {pages}
  {136406} (\bibinfo {year} {2008})}\BibitemShut {NoStop}%
\bibitem [{\citenamefont {Timrov}\ \emph {et~al.}(2018)\citenamefont {Timrov},
  \citenamefont {Marzari},\ and\ \citenamefont {Cococcioni}}]{Timrov:2018}%
  \BibitemOpen
  \bibfield  {author} {\bibinfo {author} {\bibfnamefont {I.}~\bibnamefont
  {Timrov}}, \bibinfo {author} {\bibfnamefont {N.}~\bibnamefont {Marzari}}, \
  and\ \bibinfo {author} {\bibfnamefont {M.}~\bibnamefont {Cococcioni}},\
  }\href@noop {} {\bibfield  {journal} {\bibinfo  {journal} {Phys. Rev. B}\
  }\textbf {\bibinfo {volume} {98}},\ \bibinfo {pages} {085127} (\bibinfo
  {year} {2018})}\BibitemShut {NoStop}%
\bibitem [{\citenamefont {Timrov}\ \emph {et~al.}(2021)\citenamefont {Timrov},
  \citenamefont {Marzari},\ and\ \citenamefont {Cococcioni}}]{Timrov:2021}%
  \BibitemOpen
  \bibfield  {author} {\bibinfo {author} {\bibfnamefont {I.}~\bibnamefont
  {Timrov}}, \bibinfo {author} {\bibfnamefont {N.}~\bibnamefont {Marzari}}, \
  and\ \bibinfo {author} {\bibfnamefont {M.}~\bibnamefont {Cococcioni}},\
  }\href@noop {} {\bibfield  {journal} {\bibinfo  {journal} {Phys. Rev. B}\
  }\textbf {\bibinfo {volume} {103}},\ \bibinfo {pages} {045141} (\bibinfo
  {year} {2021})}\BibitemShut {NoStop}%
\bibitem [{\citenamefont {Perdew}\ and\ \citenamefont
  {Zunger}(1981)}]{PhysRevB.23.5048}%
  \BibitemOpen
  \bibfield  {author} {\bibinfo {author} {\bibfnamefont {J.~P.}\ \bibnamefont
  {Perdew}}\ and\ \bibinfo {author} {\bibfnamefont {A.}~\bibnamefont
  {Zunger}},\ }\href {\doibase 10.1103/PhysRevB.23.5048} {\bibfield  {journal}
  {\bibinfo  {journal} {Phys. Rev. B}\ }\textbf {\bibinfo {volume} {23}},\
  \bibinfo {pages} {5048} (\bibinfo {year} {1981})}\BibitemShut {NoStop}%
\bibitem [{Note2()}]{Note2}%
  \BibitemOpen
  \bibinfo {note} {DFT simulations do not identify the small monoclinic
  distortion for MnO, see for example Ref.~\cite
  {schronCrystallineMagneticAnisotropy2012}}\BibitemShut {NoStop}%
\bibitem [{\citenamefont {Chung}\ \emph {et~al.}(2003)\citenamefont {Chung},
  \citenamefont {Paul}, \citenamefont {Balakrishnan}, \citenamefont {Lees},
  \citenamefont {Ivanov},\ and\ \citenamefont {Yethiraj}}]{PhysRevB.68.140406}%
  \BibitemOpen
  \bibfield  {author} {\bibinfo {author} {\bibfnamefont {E.~M.~L.}\
  \bibnamefont {Chung}}, \bibinfo {author} {\bibfnamefont {D.~M.}\ \bibnamefont
  {Paul}}, \bibinfo {author} {\bibfnamefont {G.}~\bibnamefont {Balakrishnan}},
  \bibinfo {author} {\bibfnamefont {M.~R.}\ \bibnamefont {Lees}}, \bibinfo
  {author} {\bibfnamefont {A.}~\bibnamefont {Ivanov}}, \ and\ \bibinfo {author}
  {\bibfnamefont {M.}~\bibnamefont {Yethiraj}},\ }\href {\doibase
  10.1103/PhysRevB.68.140406} {\bibfield  {journal} {\bibinfo  {journal} {Phys.
  Rev. B}\ }\textbf {\bibinfo {volume} {68}},\ \bibinfo {pages} {140406(R)}
  (\bibinfo {year} {2003})}\BibitemShut {NoStop}%
\bibitem [{\citenamefont {Lim}\ \emph {et~al.}(2016)\citenamefont {Lim},
  \citenamefont {Saldana-Greco},\ and\ \citenamefont
  {Rappe}}]{PhysRevB.94.165151}%
  \BibitemOpen
  \bibfield  {author} {\bibinfo {author} {\bibfnamefont {J.~S.}\ \bibnamefont
  {Lim}}, \bibinfo {author} {\bibfnamefont {D.}~\bibnamefont {Saldana-Greco}},
  \ and\ \bibinfo {author} {\bibfnamefont {A.~M.}\ \bibnamefont {Rappe}},\
  }\href {\doibase 10.1103/PhysRevB.94.165151} {\bibfield  {journal} {\bibinfo
  {journal} {Phys. Rev. B}\ }\textbf {\bibinfo {volume} {94}},\ \bibinfo
  {pages} {165151} (\bibinfo {year} {2016})}\BibitemShut {NoStop}%
\bibitem [{\citenamefont {Pask}\ \emph {et~al.}(2001)\citenamefont {Pask},
  \citenamefont {Singh}, \citenamefont {Mazin}, \citenamefont {Hellberg},\ and\
  \citenamefont {Kortus}}]{paskStructuralElectronicMagnetic2001}%
  \BibitemOpen
  \bibfield  {author} {\bibinfo {author} {\bibfnamefont {J.~E.}\ \bibnamefont
  {Pask}}, \bibinfo {author} {\bibfnamefont {D.~J.}\ \bibnamefont {Singh}},
  \bibinfo {author} {\bibfnamefont {I.~I.}\ \bibnamefont {Mazin}}, \bibinfo
  {author} {\bibfnamefont {C.~S.}\ \bibnamefont {Hellberg}}, \ and\ \bibinfo
  {author} {\bibfnamefont {J.}~\bibnamefont {Kortus}},\ }\href {\doibase
  10.1103/PhysRevB.64.024403} {\bibfield  {journal} {\bibinfo  {journal}
  {Physical Review B}\ }\textbf {\bibinfo {volume} {64}},\ \bibinfo {pages}
  {024403} (\bibinfo {year} {2001})}\BibitemShut {NoStop}%
\bibitem [{\citenamefont {Schr{\"o}n}\ \emph {et~al.}(2012)\citenamefont
  {Schr{\"o}n}, \citenamefont {R{\"o}dl},\ and\ \citenamefont
  {Bechstedt}}]{schronCrystallineMagneticAnisotropy2012}%
  \BibitemOpen
  \bibfield  {author} {\bibinfo {author} {\bibfnamefont {A.}~\bibnamefont
  {Schr{\"o}n}}, \bibinfo {author} {\bibfnamefont {C.}~\bibnamefont
  {R{\"o}dl}}, \ and\ \bibinfo {author} {\bibfnamefont {F.}~\bibnamefont
  {Bechstedt}},\ }\href {\doibase 10.1103/PhysRevB.86.115134} {\bibfield
  {journal} {\bibinfo  {journal} {Phys. Rev. B}\ }\textbf {\bibinfo {volume}
  {86}},\ \bibinfo {pages} {115134} (\bibinfo {year} {2012})}\BibitemShut
  {NoStop}%
\bibitem [{\citenamefont {Franchini}\ \emph {et~al.}(2005)\citenamefont
  {Franchini}, \citenamefont {Bayer}, \citenamefont {Podloucky}, \citenamefont
  {Paier},\ and\ \citenamefont
  {Kresse}}]{franchiniDensityFunctionalTheory2005}%
  \BibitemOpen
  \bibfield  {author} {\bibinfo {author} {\bibfnamefont {C.}~\bibnamefont
  {Franchini}}, \bibinfo {author} {\bibfnamefont {V.}~\bibnamefont {Bayer}},
  \bibinfo {author} {\bibfnamefont {R.}~\bibnamefont {Podloucky}}, \bibinfo
  {author} {\bibfnamefont {J.}~\bibnamefont {Paier}}, \ and\ \bibinfo {author}
  {\bibfnamefont {G.}~\bibnamefont {Kresse}},\ }\href {\doibase
  10.1103/PhysRevB.72.045132} {\bibfield  {journal} {\bibinfo  {journal} {Phys.
  Rev. B}\ }\textbf {\bibinfo {volume} {72}},\ \bibinfo {pages} {045132}
  (\bibinfo {year} {2005})}\BibitemShut {NoStop}%
\bibitem [{\citenamefont {Mellergård}\ \emph {et~al.}(1998)\citenamefont
  {Mellergård}, \citenamefont {McGreevy}, \citenamefont {Wannberg},\ and\
  \citenamefont {Trostell}}]{mnmoment}%
  \BibitemOpen
  \bibfield  {author} {\bibinfo {author} {\bibfnamefont {A.}~\bibnamefont
  {Mellergård}}, \bibinfo {author} {\bibfnamefont {R.~L.}\ \bibnamefont
  {McGreevy}}, \bibinfo {author} {\bibfnamefont {A.}~\bibnamefont {Wannberg}},
  \ and\ \bibinfo {author} {\bibfnamefont {B.}~\bibnamefont {Trostell}},\
  }\href {\doibase 10.1088/0953-8984/10/42/006} {\bibfield  {journal} {\bibinfo
   {journal} {Journal of Physics: Condensed Matter}\ }\textbf {\bibinfo
  {volume} {10}},\ \bibinfo {pages} {9401} (\bibinfo {year}
  {1998})}\BibitemShut {NoStop}%
\bibitem [{\citenamefont {Bonfante}\ \emph {et~al.}(1972)\citenamefont
  {Bonfante}, \citenamefont {Hennion}, \citenamefont {Moussa},\ and\
  \citenamefont {Pepy}}]{BONFANTE1972553}%
  \BibitemOpen
  \bibfield  {author} {\bibinfo {author} {\bibfnamefont {M.}~\bibnamefont
  {Bonfante}}, \bibinfo {author} {\bibfnamefont {B.}~\bibnamefont {Hennion}},
  \bibinfo {author} {\bibfnamefont {F.}~\bibnamefont {Moussa}}, \ and\ \bibinfo
  {author} {\bibfnamefont {G.}~\bibnamefont {Pepy}},\ }\href {\doibase
  https://doi.org/10.1016/0038-1098(72)90065-8} {\bibfield  {journal} {\bibinfo
   {journal} {Solid State Communications}\ }\textbf {\bibinfo {volume} {10}},\
  \bibinfo {pages} {553} (\bibinfo {year} {1972})}\BibitemShut {NoStop}%
\bibitem [{\citenamefont {Wang}\ \emph {et~al.}(2011)\citenamefont {Wang},
  \citenamefont {Baker}, \citenamefont {Lumsden}, \citenamefont {Nagler},
  \citenamefont {Heller}, \citenamefont {Baker}, \citenamefont {Deen},
  \citenamefont {Cranswick}, \citenamefont {Su},\ and\ \citenamefont
  {Christianson}}]{PhysRevB.83.214418}%
  \BibitemOpen
  \bibfield  {author} {\bibinfo {author} {\bibfnamefont {C.~H.}\ \bibnamefont
  {Wang}}, \bibinfo {author} {\bibfnamefont {S.~N.}\ \bibnamefont {Baker}},
  \bibinfo {author} {\bibfnamefont {M.~D.}\ \bibnamefont {Lumsden}}, \bibinfo
  {author} {\bibfnamefont {S.~E.}\ \bibnamefont {Nagler}}, \bibinfo {author}
  {\bibfnamefont {W.~T.}\ \bibnamefont {Heller}}, \bibinfo {author}
  {\bibfnamefont {G.~A.}\ \bibnamefont {Baker}}, \bibinfo {author}
  {\bibfnamefont {P.~D.}\ \bibnamefont {Deen}}, \bibinfo {author}
  {\bibfnamefont {L.~M.~D.}\ \bibnamefont {Cranswick}}, \bibinfo {author}
  {\bibfnamefont {Y.}~\bibnamefont {Su}}, \ and\ \bibinfo {author}
  {\bibfnamefont {A.~D.}\ \bibnamefont {Christianson}},\ }\href {\doibase
  10.1103/PhysRevB.83.214418} {\bibfield  {journal} {\bibinfo  {journal} {Phys.
  Rev. B}\ }\textbf {\bibinfo {volume} {83}},\ \bibinfo {pages} {214418}
  (\bibinfo {year} {2011})}\BibitemShut {NoStop}%
\bibitem [{\citenamefont {Huddart}\ \emph {et~al.}(2022)\citenamefont
  {Huddart}, \citenamefont {Hernández-Melián}, \citenamefont {Hicken},
  \citenamefont {Gomilšek}, \citenamefont {Hawkhead}, \citenamefont {Clark},
  \citenamefont {Pratt},\ and\ \citenamefont {Lancaster}}]{HUDDART2022108488}%
  \BibitemOpen
  \bibfield  {author} {\bibinfo {author} {\bibfnamefont {B.}~\bibnamefont
  {Huddart}}, \bibinfo {author} {\bibfnamefont {A.}~\bibnamefont
  {Hernández-Melián}}, \bibinfo {author} {\bibfnamefont {T.}~\bibnamefont
  {Hicken}}, \bibinfo {author} {\bibfnamefont {M.}~\bibnamefont {Gomilšek}},
  \bibinfo {author} {\bibfnamefont {Z.}~\bibnamefont {Hawkhead}}, \bibinfo
  {author} {\bibfnamefont {S.}~\bibnamefont {Clark}}, \bibinfo {author}
  {\bibfnamefont {F.}~\bibnamefont {Pratt}}, \ and\ \bibinfo {author}
  {\bibfnamefont {T.}~\bibnamefont {Lancaster}},\ }\href {\doibase
  https://doi.org/10.1016/j.cpc.2022.108488} {\bibfield  {journal} {\bibinfo
  {journal} {Computer Physics Communications}\ }\textbf {\bibinfo {volume}
  {280}},\ \bibinfo {pages} {108488} (\bibinfo {year} {2022})}\BibitemShut
  {NoStop}%
\bibitem [{Note3()}]{Note3}%
  \BibitemOpen
  \bibinfo {note} {We performed collinear DFT simulations, therefore the
  direction of $\protect \mathbf {S}_{e}$ cannot be established. See
  supplemental information for more details on the approximations concerning
  the evaluation of the Fermi contact field.}\BibitemShut {Stop}%
\bibitem [{\citenamefont {Onuorah}\ \emph {et~al.}(2018)\citenamefont
  {Onuorah}, \citenamefont {Bonf\`a},\ and\ \citenamefont
  {De~Renzi}}]{PhysRevB.97.174414}%
  \BibitemOpen
  \bibfield  {author} {\bibinfo {author} {\bibfnamefont {I.~J.}\ \bibnamefont
  {Onuorah}}, \bibinfo {author} {\bibfnamefont {P.}~\bibnamefont {Bonf\`a}}, \
  and\ \bibinfo {author} {\bibfnamefont {R.}~\bibnamefont {De~Renzi}},\ }\href
  {\doibase 10.1103/PhysRevB.97.174414} {\bibfield  {journal} {\bibinfo
  {journal} {Phys. Rev. B}\ }\textbf {\bibinfo {volume} {97}},\ \bibinfo
  {pages} {174414} (\bibinfo {year} {2018})}\BibitemShut {NoStop}%
\bibitem [{\citenamefont {Onuorah}\ \emph {et~al.}(2019)\citenamefont
  {Onuorah}, \citenamefont {Bonf\`a}, \citenamefont {De~Renzi}, \citenamefont
  {Monacelli}, \citenamefont {Mauri}, \citenamefont {Calandra},\ and\
  \citenamefont {Errea}}]{PhysRevMaterials.3.073804}%
  \BibitemOpen
  \bibfield  {author} {\bibinfo {author} {\bibfnamefont {I.~J.}\ \bibnamefont
  {Onuorah}}, \bibinfo {author} {\bibfnamefont {P.}~\bibnamefont {Bonf\`a}},
  \bibinfo {author} {\bibfnamefont {R.}~\bibnamefont {De~Renzi}}, \bibinfo
  {author} {\bibfnamefont {L.}~\bibnamefont {Monacelli}}, \bibinfo {author}
  {\bibfnamefont {F.}~\bibnamefont {Mauri}}, \bibinfo {author} {\bibfnamefont
  {M.}~\bibnamefont {Calandra}}, \ and\ \bibinfo {author} {\bibfnamefont
  {I.}~\bibnamefont {Errea}},\ }\href {\doibase
  10.1103/PhysRevMaterials.3.073804} {\bibfield  {journal} {\bibinfo  {journal}
  {Phys. Rev. Mater.}\ }\textbf {\bibinfo {volume} {3}},\ \bibinfo {pages}
  {073804} (\bibinfo {year} {2019})}\BibitemShut {NoStop}%
\bibitem [{\citenamefont {Frandsen}\ and\ \citenamefont
  {Billinge}(2015)}]{Frandsen:vk5003}%
  \BibitemOpen
  \bibfield  {author} {\bibinfo {author} {\bibfnamefont {B.~A.}\ \bibnamefont
  {Frandsen}}\ and\ \bibinfo {author} {\bibfnamefont {S.~J.~L.}\ \bibnamefont
  {Billinge}},\ }\href {\doibase 10.1107/S205327331500306X} {\bibfield
  {journal} {\bibinfo  {journal} {Acta Crystallographica Section A}\ }\textbf
  {\bibinfo {volume} {71}},\ \bibinfo {pages} {325} (\bibinfo {year}
  {2015})}\BibitemShut {NoStop}%
\bibitem [{\citenamefont {Massidda}\ \emph {et~al.}(1999)\citenamefont
  {Massidda}, \citenamefont {Posternak}, \citenamefont {Baldereschi},\ and\
  \citenamefont {Resta}}]{PhysRevLett.82.430}%
  \BibitemOpen
  \bibfield  {author} {\bibinfo {author} {\bibfnamefont {S.}~\bibnamefont
  {Massidda}}, \bibinfo {author} {\bibfnamefont {M.}~\bibnamefont {Posternak}},
  \bibinfo {author} {\bibfnamefont {A.}~\bibnamefont {Baldereschi}}, \ and\
  \bibinfo {author} {\bibfnamefont {R.}~\bibnamefont {Resta}},\ }\href
  {\doibase 10.1103/PhysRevLett.82.430} {\bibfield  {journal} {\bibinfo
  {journal} {Phys. Rev. Lett.}\ }\textbf {\bibinfo {volume} {82}},\ \bibinfo
  {pages} {430} (\bibinfo {year} {1999})}\BibitemShut {NoStop}%
\bibitem [{\citenamefont {Jauch}\ and\ \citenamefont
  {Reehuis}(2003)}]{PhysRevB.67.184420}%
  \BibitemOpen
  \bibfield  {author} {\bibinfo {author} {\bibfnamefont {W.}~\bibnamefont
  {Jauch}}\ and\ \bibinfo {author} {\bibfnamefont {M.}~\bibnamefont
  {Reehuis}},\ }\href {\doibase 10.1103/PhysRevB.67.184420} {\bibfield
  {journal} {\bibinfo  {journal} {Phys. Rev. B}\ }\textbf {\bibinfo {volume}
  {67}},\ \bibinfo {pages} {184420} (\bibinfo {year} {2003})}\BibitemShut
  {NoStop}%
\bibitem [{\citenamefont {Vidal}\ \emph {et~al.}(2002)\citenamefont {Vidal},
  \citenamefont {Vidal-Valat}, \citenamefont {Kurki-Suonio},\ and\
  \citenamefont {Kurki-Suonio}}]{Vidal2002}%
  \BibitemOpen
  \bibfield  {author} {\bibinfo {author} {\bibfnamefont {J.-P.}\ \bibnamefont
  {Vidal}}, \bibinfo {author} {\bibfnamefont {G.}~\bibnamefont {Vidal-Valat}},
  \bibinfo {author} {\bibfnamefont {K.}~\bibnamefont {Kurki-Suonio}}, \ and\
  \bibinfo {author} {\bibfnamefont {R.}~\bibnamefont {Kurki-Suonio}},\ }\href
  {\doibase 10.1134/1.1481915} {\bibfield  {journal} {\bibinfo  {journal}
  {Crystallogr. Rep.}\ }\textbf {\bibinfo {volume} {47}},\ \bibinfo {pages}
  {347} (\bibinfo {year} {2002})}\BibitemShut {NoStop}%
\bibitem [{Note4()}]{Note4}%
  \BibitemOpen
  \bibinfo {note} {In a spin unpolarized simulation the energy difference
  vanishes.}\BibitemShut {Stop}%
\bibitem [{\citenamefont {Lee}\ \emph {et~al.}(2016)\citenamefont {Lee},
  \citenamefont {Ishikawa}, \citenamefont {Miao}, \citenamefont {Torii},
  \citenamefont {Ishigaki},\ and\ \citenamefont {Kamiyama}}]{Lee2016}%
  \BibitemOpen
  \bibfield  {author} {\bibinfo {author} {\bibfnamefont {S.}~\bibnamefont
  {Lee}}, \bibinfo {author} {\bibfnamefont {Y.}~\bibnamefont {Ishikawa}},
  \bibinfo {author} {\bibfnamefont {P.}~\bibnamefont {Miao}}, \bibinfo {author}
  {\bibfnamefont {S.}~\bibnamefont {Torii}}, \bibinfo {author} {\bibfnamefont
  {T.}~\bibnamefont {Ishigaki}}, \ and\ \bibinfo {author} {\bibfnamefont
  {T.}~\bibnamefont {Kamiyama}},\ }\href {\doibase 10.1103/PhysRevB.93.064429}
  {\bibfield  {journal} {\bibinfo  {journal} {Phys. Rev. B}\ }\textbf {\bibinfo
  {volume} {93}},\ \bibinfo {pages} {064429} (\bibinfo {year}
  {2016})}\BibitemShut {NoStop}%
\bibitem [{\citenamefont {Kofstad}(1983)}]{KOFSTAD1983879}%
  \BibitemOpen
  \bibfield  {author} {\bibinfo {author} {\bibfnamefont {P.}~\bibnamefont
  {Kofstad}},\ }\href {\doibase https://doi.org/10.1016/0022-3697(83)90126-9}
  {\bibfield  {journal} {\bibinfo  {journal} {Journal of Physics and Chemistry
  of Solids}\ }\textbf {\bibinfo {volume} {44}},\ \bibinfo {pages} {879}
  (\bibinfo {year} {1983})}\BibitemShut {NoStop}%
\bibitem [{\citenamefont {Hed}\ and\ \citenamefont
  {Tannhauser}(2004)}]{10.1063/1.1712241}%
  \BibitemOpen
  \bibfield  {author} {\bibinfo {author} {\bibfnamefont {A.~Z.}\ \bibnamefont
  {Hed}}\ and\ \bibinfo {author} {\bibfnamefont {D.~S.}\ \bibnamefont
  {Tannhauser}},\ }\href {\doibase 10.1063/1.1712241} {\bibfield  {journal}
  {\bibinfo  {journal} {The Journal of Chemical Physics}\ }\textbf {\bibinfo
  {volume} {47}},\ \bibinfo {pages} {2090} (\bibinfo {year}
  {2004})}\BibitemShut {NoStop}%
\bibitem [{\citenamefont {Bitsanis}\ \emph {et~al.}(1987)\citenamefont
  {Bitsanis}, \citenamefont {Tirrell},\ and\ \citenamefont
  {Davis}}]{PhysRevA.36.958}%
  \BibitemOpen
  \bibfield  {author} {\bibinfo {author} {\bibfnamefont {I.}~\bibnamefont
  {Bitsanis}}, \bibinfo {author} {\bibfnamefont {M.}~\bibnamefont {Tirrell}}, \
  and\ \bibinfo {author} {\bibfnamefont {H.~T.}\ \bibnamefont {Davis}},\ }\href
  {\doibase 10.1103/PhysRevA.36.958} {\bibfield  {journal} {\bibinfo  {journal}
  {Phys. Rev. A}\ }\textbf {\bibinfo {volume} {36}},\ \bibinfo {pages} {958}
  (\bibinfo {year} {1987})}\BibitemShut {NoStop}%
\bibitem [{\citenamefont {Mattoni}\ \emph {et~al.}(2015)\citenamefont
  {Mattoni}, \citenamefont {Filippetti}, \citenamefont {Saba},\ and\
  \citenamefont {Delugas}}]{mattoniMethylammoniumRotationalDynamics2015}%
  \BibitemOpen
  \bibfield  {author} {\bibinfo {author} {\bibfnamefont {A.}~\bibnamefont
  {Mattoni}}, \bibinfo {author} {\bibfnamefont {A.}~\bibnamefont {Filippetti}},
  \bibinfo {author} {\bibfnamefont {M.~I.}\ \bibnamefont {Saba}}, \ and\
  \bibinfo {author} {\bibfnamefont {P.}~\bibnamefont {Delugas}},\ }\href
  {\doibase 10.1021/acs.jpcc.5b04283} {\bibfield  {journal} {\bibinfo
  {journal} {The Journal of Physical Chemistry C}\ }\textbf {\bibinfo {volume}
  {119}},\ \bibinfo {pages} {17421} (\bibinfo {year} {2015})}\BibitemShut
  {NoStop}%
\bibitem [{\citenamefont {Zwanzig}\ and\ \citenamefont
  {Ailawadi}(1969)}]{PhysRev.182.280}%
  \BibitemOpen
  \bibfield  {author} {\bibinfo {author} {\bibfnamefont {R.}~\bibnamefont
  {Zwanzig}}\ and\ \bibinfo {author} {\bibfnamefont {N.~K.}\ \bibnamefont
  {Ailawadi}},\ }\href {\doibase 10.1103/PhysRev.182.280} {\bibfield  {journal}
  {\bibinfo  {journal} {Phys. Rev.}\ }\textbf {\bibinfo {volume} {182}},\
  \bibinfo {pages} {280} (\bibinfo {year} {1969})}\BibitemShut {NoStop}%
\bibitem [{\citenamefont {He}\ \emph {et~al.}(2018)\citenamefont {He},
  \citenamefont {Zhu}, \citenamefont {Epstein},\ and\ \citenamefont
  {Mo}}]{He2018}%
  \BibitemOpen
  \bibfield  {author} {\bibinfo {author} {\bibfnamefont {X.}~\bibnamefont
  {He}}, \bibinfo {author} {\bibfnamefont {Y.}~\bibnamefont {Zhu}}, \bibinfo
  {author} {\bibfnamefont {A.}~\bibnamefont {Epstein}}, \ and\ \bibinfo
  {author} {\bibfnamefont {Y.}~\bibnamefont {Mo}},\ }\href {\doibase
  10.1038/s41524-018-0074-y} {\bibfield  {journal} {\bibinfo  {journal} {npj
  Computational Materials}\ }\textbf {\bibinfo {volume} {4}},\ \bibinfo {pages}
  {18} (\bibinfo {year} {2018})}\BibitemShut {NoStop}%
\bibitem [{Note5()}]{Note5}%
  \BibitemOpen
  \bibinfo {note} {The experimental estimate includes a correction due to the
  electronic correlation time, see cited reference for details.}\BibitemShut
  {Stop}%
\bibitem [{\citenamefont {Nishiyama}\ \emph {et~al.}(2001)\citenamefont
  {Nishiyama}, \citenamefont {Higemoto}, \citenamefont {Shimomura},
  \citenamefont {Koda}, \citenamefont {Maruta}, \citenamefont {Nishiyama},\
  and\ \citenamefont {Zheng}}]{Nishiyama2001}%
  \BibitemOpen
  \bibfield  {author} {\bibinfo {author} {\bibfnamefont {K.}~\bibnamefont
  {Nishiyama}}, \bibinfo {author} {\bibfnamefont {W.}~\bibnamefont {Higemoto}},
  \bibinfo {author} {\bibfnamefont {K.}~\bibnamefont {Shimomura}}, \bibinfo
  {author} {\bibfnamefont {A.}~\bibnamefont {Koda}}, \bibinfo {author}
  {\bibfnamefont {G.}~\bibnamefont {Maruta}}, \bibinfo {author} {\bibfnamefont
  {S.~W.}\ \bibnamefont {Nishiyama}}, \ and\ \bibinfo {author} {\bibfnamefont
  {X.~G.}\ \bibnamefont {Zheng}},\ }\href {\doibase 10.1023/A:1020504620851}
  {\bibfield  {journal} {\bibinfo  {journal} {Hyperfine Interact}\ }\textbf
  {\bibinfo {volume} {136}},\ \bibinfo {pages} {289} (\bibinfo {year}
  {2001})}\BibitemShut {NoStop}%
\bibitem [{\citenamefont {Nishiyama}\ \emph {et~al.}(1997)\citenamefont
  {Nishiyama}, \citenamefont {Ohira}, \citenamefont {Dawson},\ and\
  \citenamefont {Higemoto}}]{Nishiyama1997}%
  \BibitemOpen
  \bibfield  {author} {\bibinfo {author} {\bibfnamefont {K.}~\bibnamefont
  {Nishiyama}}, \bibinfo {author} {\bibfnamefont {S.}~\bibnamefont {Ohira}},
  \bibinfo {author} {\bibfnamefont {W.}~\bibnamefont {Dawson}}, \ and\ \bibinfo
  {author} {\bibfnamefont {W.}~\bibnamefont {Higemoto}},\ }\href {\doibase
  10.1023/A:1012681305657} {\bibfield  {journal} {\bibinfo  {journal}
  {Hyperfine Interact}\ }\textbf {\bibinfo {volume} {104}},\ \bibinfo {pages}
  {349} (\bibinfo {year} {1997})}\BibitemShut {NoStop}%
\bibitem [{\citenamefont {Dehn}\ \emph {et~al.}(2021)\citenamefont {Dehn},
  \citenamefont {Shenton}, \citenamefont {Arseneau}, \citenamefont
  {MacFarlane}, \citenamefont {Morris}, \citenamefont {Maign\'e}, \citenamefont
  {Spaldin},\ and\ \citenamefont {Kiefl}}]{PhysRevLett.126.037202}%
  \BibitemOpen
  \bibfield  {author} {\bibinfo {author} {\bibfnamefont {M.~H.}\ \bibnamefont
  {Dehn}}, \bibinfo {author} {\bibfnamefont {J.~K.}\ \bibnamefont {Shenton}},
  \bibinfo {author} {\bibfnamefont {D.~J.}\ \bibnamefont {Arseneau}}, \bibinfo
  {author} {\bibfnamefont {W.~A.}\ \bibnamefont {MacFarlane}}, \bibinfo
  {author} {\bibfnamefont {G.~D.}\ \bibnamefont {Morris}}, \bibinfo {author}
  {\bibfnamefont {A.}~\bibnamefont {Maign\'e}}, \bibinfo {author}
  {\bibfnamefont {N.~A.}\ \bibnamefont {Spaldin}}, \ and\ \bibinfo {author}
  {\bibfnamefont {R.~F.}\ \bibnamefont {Kiefl}},\ }\href {\doibase
  10.1103/PhysRevLett.126.037202} {\bibfield  {journal} {\bibinfo  {journal}
  {Phys. Rev. Lett.}\ }\textbf {\bibinfo {volume} {126}},\ \bibinfo {pages}
  {037202} (\bibinfo {year} {2021})}\BibitemShut {NoStop}%
\bibitem [{\citenamefont {Dehn}\ \emph {et~al.}(2020)\citenamefont {Dehn},
  \citenamefont {Shenton}, \citenamefont {Holenstein}, \citenamefont {Meier},
  \citenamefont {Arseneau}, \citenamefont {Cortie}, \citenamefont {Hitti},
  \citenamefont {Fang}, \citenamefont {MacFarlane}, \citenamefont {McFadden},
  \citenamefont {Morris}, \citenamefont {Salman}, \citenamefont {Luetkens},
  \citenamefont {Spaldin}, \citenamefont {Fechner},\ and\ \citenamefont
  {Kiefl}}]{PhysRevX.10.011036}%
  \BibitemOpen
  \bibfield  {author} {\bibinfo {author} {\bibfnamefont {M.~H.}\ \bibnamefont
  {Dehn}}, \bibinfo {author} {\bibfnamefont {J.~K.}\ \bibnamefont {Shenton}},
  \bibinfo {author} {\bibfnamefont {S.}~\bibnamefont {Holenstein}}, \bibinfo
  {author} {\bibfnamefont {Q.~N.}\ \bibnamefont {Meier}}, \bibinfo {author}
  {\bibfnamefont {D.~J.}\ \bibnamefont {Arseneau}}, \bibinfo {author}
  {\bibfnamefont {D.~L.}\ \bibnamefont {Cortie}}, \bibinfo {author}
  {\bibfnamefont {B.}~\bibnamefont {Hitti}}, \bibinfo {author} {\bibfnamefont
  {A.~C.~Y.}\ \bibnamefont {Fang}}, \bibinfo {author} {\bibfnamefont {W.~A.}\
  \bibnamefont {MacFarlane}}, \bibinfo {author} {\bibfnamefont {R.~M.~L.}\
  \bibnamefont {McFadden}}, \bibinfo {author} {\bibfnamefont {G.~D.}\
  \bibnamefont {Morris}}, \bibinfo {author} {\bibfnamefont {Z.}~\bibnamefont
  {Salman}}, \bibinfo {author} {\bibfnamefont {H.}~\bibnamefont {Luetkens}},
  \bibinfo {author} {\bibfnamefont {N.~A.}\ \bibnamefont {Spaldin}}, \bibinfo
  {author} {\bibfnamefont {M.}~\bibnamefont {Fechner}}, \ and\ \bibinfo
  {author} {\bibfnamefont {R.~F.}\ \bibnamefont {Kiefl}},\ }\href {\doibase
  10.1103/PhysRevX.10.011036} {\bibfield  {journal} {\bibinfo  {journal} {Phys.
  Rev. X}\ }\textbf {\bibinfo {volume} {10}},\ \bibinfo {pages} {011036}
  (\bibinfo {year} {2020})}\BibitemShut {NoStop}%
\bibitem [{\citenamefont {García}\ \emph {et~al.}(2020)\citenamefont
  {García}, \citenamefont {Papior}, \citenamefont {Akhtar}, \citenamefont
  {Artacho}, \citenamefont {Blum}, \citenamefont {Bosoni}, \citenamefont
  {Brandimarte}, \citenamefont {Brandbyge}, \citenamefont {Cerdá},
  \citenamefont {Corsetti}, \citenamefont {Cuadrado}, \citenamefont {Dikan},
  \citenamefont {Ferrer}, \citenamefont {Gale}, \citenamefont
  {García-Fernández}, \citenamefont {García-Suárez}, \citenamefont
  {García}, \citenamefont {Huhs}, \citenamefont {Illera}, \citenamefont
  {Korytár}, \citenamefont {Koval}, \citenamefont {Lebedeva}, \citenamefont
  {Lin}, \citenamefont {López-Tarifa}, \citenamefont {Mayo}, \citenamefont
  {Mohr}, \citenamefont {Ordejón}, \citenamefont {Postnikov}, \citenamefont
  {Pouillon}, \citenamefont {Pruneda}, \citenamefont {Robles}, \citenamefont
  {Sánchez-Portal}, \citenamefont {Soler}, \citenamefont {Ullah},
  \citenamefont {Yu},\ and\ \citenamefont {Junquera}}]{10.1063/5.0005077}%
  \BibitemOpen
  \bibfield  {author} {\bibinfo {author} {\bibfnamefont {A.}~\bibnamefont
  {García}}, \bibinfo {author} {\bibfnamefont {N.}~\bibnamefont {Papior}},
  \bibinfo {author} {\bibfnamefont {A.}~\bibnamefont {Akhtar}}, \bibinfo
  {author} {\bibfnamefont {E.}~\bibnamefont {Artacho}}, \bibinfo {author}
  {\bibfnamefont {V.}~\bibnamefont {Blum}}, \bibinfo {author} {\bibfnamefont
  {E.}~\bibnamefont {Bosoni}}, \bibinfo {author} {\bibfnamefont
  {P.}~\bibnamefont {Brandimarte}}, \bibinfo {author} {\bibfnamefont
  {M.}~\bibnamefont {Brandbyge}}, \bibinfo {author} {\bibfnamefont {J.~I.}\
  \bibnamefont {Cerdá}}, \bibinfo {author} {\bibfnamefont {F.}~\bibnamefont
  {Corsetti}}, \bibinfo {author} {\bibfnamefont {R.}~\bibnamefont {Cuadrado}},
  \bibinfo {author} {\bibfnamefont {V.}~\bibnamefont {Dikan}}, \bibinfo
  {author} {\bibfnamefont {J.}~\bibnamefont {Ferrer}}, \bibinfo {author}
  {\bibfnamefont {J.}~\bibnamefont {Gale}}, \bibinfo {author} {\bibfnamefont
  {P.}~\bibnamefont {García-Fernández}}, \bibinfo {author} {\bibfnamefont
  {V.~M.}\ \bibnamefont {García-Suárez}}, \bibinfo {author} {\bibfnamefont
  {S.}~\bibnamefont {García}}, \bibinfo {author} {\bibfnamefont
  {G.}~\bibnamefont {Huhs}}, \bibinfo {author} {\bibfnamefont {S.}~\bibnamefont
  {Illera}}, \bibinfo {author} {\bibfnamefont {R.}~\bibnamefont {Korytár}},
  \bibinfo {author} {\bibfnamefont {P.}~\bibnamefont {Koval}}, \bibinfo
  {author} {\bibfnamefont {I.}~\bibnamefont {Lebedeva}}, \bibinfo {author}
  {\bibfnamefont {L.}~\bibnamefont {Lin}}, \bibinfo {author} {\bibfnamefont
  {P.}~\bibnamefont {López-Tarifa}}, \bibinfo {author} {\bibfnamefont {S.~G.}\
  \bibnamefont {Mayo}}, \bibinfo {author} {\bibfnamefont {S.}~\bibnamefont
  {Mohr}}, \bibinfo {author} {\bibfnamefont {P.}~\bibnamefont {Ordejón}},
  \bibinfo {author} {\bibfnamefont {A.}~\bibnamefont {Postnikov}}, \bibinfo
  {author} {\bibfnamefont {Y.}~\bibnamefont {Pouillon}}, \bibinfo {author}
  {\bibfnamefont {M.}~\bibnamefont {Pruneda}}, \bibinfo {author} {\bibfnamefont
  {R.}~\bibnamefont {Robles}}, \bibinfo {author} {\bibfnamefont
  {D.}~\bibnamefont {Sánchez-Portal}}, \bibinfo {author} {\bibfnamefont
  {J.~M.}\ \bibnamefont {Soler}}, \bibinfo {author} {\bibfnamefont
  {R.}~\bibnamefont {Ullah}}, \bibinfo {author} {\bibfnamefont {V.~W.-z.}\
  \bibnamefont {Yu}}, \ and\ \bibinfo {author} {\bibfnamefont {J.}~\bibnamefont
  {Junquera}},\ }\href {\doibase 10.1063/5.0005077} {\bibfield  {journal}
  {\bibinfo  {journal} {The Journal of Chemical Physics}\ }\textbf {\bibinfo
  {volume} {152}},\ \bibinfo {pages} {204108} (\bibinfo {year}
  {2020})}\BibitemShut {NoStop}%
\bibitem [{\citenamefont {Baroni}\ \emph {et~al.}(2001)\citenamefont {Baroni},
  \citenamefont {de~Gironcoli}, \citenamefont {Dal~Corso},\ and\ \citenamefont
  {Giannozzi}}]{Baroni:2001}%
  \BibitemOpen
  \bibfield  {author} {\bibinfo {author} {\bibfnamefont {S.}~\bibnamefont
  {Baroni}}, \bibinfo {author} {\bibfnamefont {S.}~\bibnamefont
  {de~Gironcoli}}, \bibinfo {author} {\bibfnamefont {A.}~\bibnamefont
  {Dal~Corso}}, \ and\ \bibinfo {author} {\bibfnamefont {P.}~\bibnamefont
  {Giannozzi}},\ }\href@noop {} {\bibfield  {journal} {\bibinfo  {journal}
  {Rev. Mod. Phys.}\ }\textbf {\bibinfo {volume} {73}},\ \bibinfo {pages} {515}
  (\bibinfo {year} {2001})}\BibitemShut {NoStop}%
\bibitem [{\citenamefont {Bennett}\ \emph {et~al.}(2019)\citenamefont
  {Bennett}, \citenamefont {Hudson}, \citenamefont {Metz}, \citenamefont
  {Liang}, \citenamefont {Spurgeon}, \citenamefont {Cui},\ and\ \citenamefont
  {Mason}}]{BENNETT2019109137}%
  \BibitemOpen
  \bibfield  {author} {\bibinfo {author} {\bibfnamefont {J.~W.}\ \bibnamefont
  {Bennett}}, \bibinfo {author} {\bibfnamefont {B.~G.}\ \bibnamefont {Hudson}},
  \bibinfo {author} {\bibfnamefont {I.~K.}\ \bibnamefont {Metz}}, \bibinfo
  {author} {\bibfnamefont {D.}~\bibnamefont {Liang}}, \bibinfo {author}
  {\bibfnamefont {S.}~\bibnamefont {Spurgeon}}, \bibinfo {author}
  {\bibfnamefont {Q.}~\bibnamefont {Cui}}, \ and\ \bibinfo {author}
  {\bibfnamefont {S.~E.}\ \bibnamefont {Mason}},\ }\href {\doibase
  https://doi.org/10.1016/j.commatsci.2019.109137} {\bibfield  {journal}
  {\bibinfo  {journal} {Computational Materials Science}\ }\textbf {\bibinfo
  {volume} {170}},\ \bibinfo {pages} {109137} (\bibinfo {year}
  {2019})}\BibitemShut {NoStop}%
\bibitem [{\citenamefont {Bonf{\`{a}}}\ and\ \citenamefont {{De
  Renzi}}(2016)}]{Bonfa2016}%
  \BibitemOpen
  \bibfield  {author} {\bibinfo {author} {\bibfnamefont {P.}~\bibnamefont
  {Bonf{\`{a}}}}\ and\ \bibinfo {author} {\bibfnamefont {R.}~\bibnamefont {{De
  Renzi}}},\ }\href {\doibase 10.7566/JPSJ.85.091014} {\bibfield  {journal}
  {\bibinfo  {journal} {J. Phys. Soc. Jpn.}\ }\textbf {\bibinfo {volume}
  {85}},\ \bibinfo {pages} {091014} (\bibinfo {year} {2016})}\BibitemShut
  {NoStop}%
\bibitem [{\citenamefont {Chmiela}\ \emph {et~al.}(2023)\citenamefont
  {Chmiela}, \citenamefont {Vassilev-Galindo}, \citenamefont {Unke},
  \citenamefont {Kabylda}, \citenamefont {Sauceda}, \citenamefont
  {Tkatchenko},\ and\ \citenamefont {M\"{u}ller}}]{chmiela2023}%
  \BibitemOpen
  \bibfield  {author} {\bibinfo {author} {\bibfnamefont {S.}~\bibnamefont
  {Chmiela}}, \bibinfo {author} {\bibfnamefont {V.}~\bibnamefont
  {Vassilev-Galindo}}, \bibinfo {author} {\bibfnamefont {O.~T.}\ \bibnamefont
  {Unke}}, \bibinfo {author} {\bibfnamefont {A.}~\bibnamefont {Kabylda}},
  \bibinfo {author} {\bibfnamefont {H.~E.}\ \bibnamefont {Sauceda}}, \bibinfo
  {author} {\bibfnamefont {A.}~\bibnamefont {Tkatchenko}}, \ and\ \bibinfo
  {author} {\bibfnamefont {K.-R.}\ \bibnamefont {M\"{u}ller}},\ }\href
  {\doibase 10.1126/sciadv.adf0873} {\bibfield  {journal} {\bibinfo  {journal}
  {Science Advances}\ }\textbf {\bibinfo {volume} {9}},\ \bibinfo {pages}
  {eadf0873} (\bibinfo {year} {2023})}\BibitemShut {NoStop}%
\bibitem [{\citenamefont {Cococcioni}\ and\ \citenamefont
  {de~Gironcoli}(2005)}]{Cococcioni:2005}%
  \BibitemOpen
  \bibfield  {author} {\bibinfo {author} {\bibfnamefont {M.}~\bibnamefont
  {Cococcioni}}\ and\ \bibinfo {author} {\bibfnamefont {S.}~\bibnamefont
  {de~Gironcoli}},\ }\href@noop {} {\bibfield  {journal} {\bibinfo  {journal}
  {Phys. Rev. B}\ }\textbf {\bibinfo {volume} {71}},\ \bibinfo {pages} {035105}
  (\bibinfo {year} {2005})}\BibitemShut {NoStop}%
\bibitem [{\citenamefont {{Dal Corso}}(2014)}]{DALCORSO2014337}%
  \BibitemOpen
  \bibfield  {author} {\bibinfo {author} {\bibfnamefont {A.}~\bibnamefont {{Dal
  Corso}}},\ }\href {\doibase https://doi.org/10.1016/j.commatsci.2014.07.043}
  {\bibfield  {journal} {\bibinfo  {journal} {Computational Materials Science}\
  }\textbf {\bibinfo {volume} {95}},\ \bibinfo {pages} {337} (\bibinfo {year}
  {2014})}\BibitemShut {NoStop}%
\bibitem [{\citenamefont {Garrity}\ \emph {et~al.}(2014)\citenamefont
  {Garrity}, \citenamefont {Bennett}, \citenamefont {Rabe},\ and\ \citenamefont
  {Vanderbilt}}]{GARRITY2014446}%
  \BibitemOpen
  \bibfield  {author} {\bibinfo {author} {\bibfnamefont {K.~F.}\ \bibnamefont
  {Garrity}}, \bibinfo {author} {\bibfnamefont {J.~W.}\ \bibnamefont
  {Bennett}}, \bibinfo {author} {\bibfnamefont {K.~M.}\ \bibnamefont {Rabe}}, \
  and\ \bibinfo {author} {\bibfnamefont {D.}~\bibnamefont {Vanderbilt}},\
  }\href {\doibase https://doi.org/10.1016/j.commatsci.2013.08.053} {\bibfield
  {journal} {\bibinfo  {journal} {Computational Materials Science}\ }\textbf
  {\bibinfo {volume} {81}},\ \bibinfo {pages} {446} (\bibinfo {year}
  {2014})}\BibitemShut {NoStop}%
\bibitem [{\citenamefont {Giannozzi}\ \emph {et~al.}(2020)\citenamefont
  {Giannozzi}, \citenamefont {Baseggio}, \citenamefont {Bonf{\`a}},
  \citenamefont {Brunato}, \citenamefont {Car}, \citenamefont {Carnimeo},
  \citenamefont {Cavazzoni}, \citenamefont {De~Gironcoli}, \citenamefont
  {Delugas}, \citenamefont {Ferrari~Ruffino} \emph
  {et~al.}}]{giannozzi2020quantum}%
  \BibitemOpen
  \bibfield  {author} {\bibinfo {author} {\bibfnamefont {P.}~\bibnamefont
  {Giannozzi}}, \bibinfo {author} {\bibfnamefont {O.}~\bibnamefont {Baseggio}},
  \bibinfo {author} {\bibfnamefont {P.}~\bibnamefont {Bonf{\`a}}}, \bibinfo
  {author} {\bibfnamefont {D.}~\bibnamefont {Brunato}}, \bibinfo {author}
  {\bibfnamefont {R.}~\bibnamefont {Car}}, \bibinfo {author} {\bibfnamefont
  {I.}~\bibnamefont {Carnimeo}}, \bibinfo {author} {\bibfnamefont
  {C.}~\bibnamefont {Cavazzoni}}, \bibinfo {author} {\bibfnamefont
  {S.}~\bibnamefont {De~Gironcoli}}, \bibinfo {author} {\bibfnamefont
  {P.}~\bibnamefont {Delugas}}, \bibinfo {author} {\bibfnamefont
  {F.}~\bibnamefont {Ferrari~Ruffino}},  \emph {et~al.},\ }\href@noop {}
  {\bibfield  {journal} {\bibinfo  {journal} {The Journal of chemical physics}\
  }\textbf {\bibinfo {volume} {152}},\ \bibinfo {pages} {154105} (\bibinfo
  {year} {2020})}\BibitemShut {NoStop}%
\bibitem [{\citenamefont {Giannozzi}\ \emph {et~al.}(2009)\citenamefont
  {Giannozzi}, \citenamefont {Baroni}, \citenamefont {Bonini}, \citenamefont
  {Calandra}, \citenamefont {Car}, \citenamefont {Cavazzoni}, \citenamefont
  {Ceresoli}, \citenamefont {Chiarotti}, \citenamefont {Cococcioni},
  \citenamefont {Dabo}, \citenamefont {Corso}, \citenamefont {de~Gironcoli},
  \citenamefont {Fabris}, \citenamefont {Fratesi}, \citenamefont {Gebauer},
  \citenamefont {Gerstmann}, \citenamefont {Gougoussis}, \citenamefont
  {Kokalj}, \citenamefont {Lazzeri}, \citenamefont {Martin-Samos},
  \citenamefont {Marzari}, \citenamefont {Mauri}, \citenamefont {Mazzarello},
  \citenamefont {Paolini}, \citenamefont {Pasquarello}, \citenamefont
  {Paulatto}, \citenamefont {Sbraccia}, \citenamefont {Scandolo}, \citenamefont
  {Sclauzero}, \citenamefont {Seitsonen}, \citenamefont {Smogunov},
  \citenamefont {Umari},\ and\ \citenamefont {Wentzcovitch}}]{Giannozzi_2009}%
  \BibitemOpen
  \bibfield  {author} {\bibinfo {author} {\bibfnamefont {P.}~\bibnamefont
  {Giannozzi}}, \bibinfo {author} {\bibfnamefont {S.}~\bibnamefont {Baroni}},
  \bibinfo {author} {\bibfnamefont {N.}~\bibnamefont {Bonini}}, \bibinfo
  {author} {\bibfnamefont {M.}~\bibnamefont {Calandra}}, \bibinfo {author}
  {\bibfnamefont {R.}~\bibnamefont {Car}}, \bibinfo {author} {\bibfnamefont
  {C.}~\bibnamefont {Cavazzoni}}, \bibinfo {author} {\bibfnamefont
  {D.}~\bibnamefont {Ceresoli}}, \bibinfo {author} {\bibfnamefont {G.~L.}\
  \bibnamefont {Chiarotti}}, \bibinfo {author} {\bibfnamefont {M.}~\bibnamefont
  {Cococcioni}}, \bibinfo {author} {\bibfnamefont {I.}~\bibnamefont {Dabo}},
  \bibinfo {author} {\bibfnamefont {A.~D.}\ \bibnamefont {Corso}}, \bibinfo
  {author} {\bibfnamefont {S.}~\bibnamefont {de~Gironcoli}}, \bibinfo {author}
  {\bibfnamefont {S.}~\bibnamefont {Fabris}}, \bibinfo {author} {\bibfnamefont
  {G.}~\bibnamefont {Fratesi}}, \bibinfo {author} {\bibfnamefont
  {R.}~\bibnamefont {Gebauer}}, \bibinfo {author} {\bibfnamefont
  {U.}~\bibnamefont {Gerstmann}}, \bibinfo {author} {\bibfnamefont
  {C.}~\bibnamefont {Gougoussis}}, \bibinfo {author} {\bibfnamefont
  {A.}~\bibnamefont {Kokalj}}, \bibinfo {author} {\bibfnamefont
  {M.}~\bibnamefont {Lazzeri}}, \bibinfo {author} {\bibfnamefont
  {L.}~\bibnamefont {Martin-Samos}}, \bibinfo {author} {\bibfnamefont
  {N.}~\bibnamefont {Marzari}}, \bibinfo {author} {\bibfnamefont
  {F.}~\bibnamefont {Mauri}}, \bibinfo {author} {\bibfnamefont
  {R.}~\bibnamefont {Mazzarello}}, \bibinfo {author} {\bibfnamefont
  {S.}~\bibnamefont {Paolini}}, \bibinfo {author} {\bibfnamefont
  {A.}~\bibnamefont {Pasquarello}}, \bibinfo {author} {\bibfnamefont
  {L.}~\bibnamefont {Paulatto}}, \bibinfo {author} {\bibfnamefont
  {C.}~\bibnamefont {Sbraccia}}, \bibinfo {author} {\bibfnamefont
  {S.}~\bibnamefont {Scandolo}}, \bibinfo {author} {\bibfnamefont
  {G.}~\bibnamefont {Sclauzero}}, \bibinfo {author} {\bibfnamefont {A.~P.}\
  \bibnamefont {Seitsonen}}, \bibinfo {author} {\bibfnamefont {A.}~\bibnamefont
  {Smogunov}}, \bibinfo {author} {\bibfnamefont {P.}~\bibnamefont {Umari}}, \
  and\ \bibinfo {author} {\bibfnamefont {R.~M.}\ \bibnamefont {Wentzcovitch}},\
  }\href {\doibase 10.1088/0953-8984/21/39/395502} {\bibfield  {journal}
  {\bibinfo  {journal} {Journal of Physics: Condensed Matter}\ }\textbf
  {\bibinfo {volume} {21}},\ \bibinfo {pages} {395502} (\bibinfo {year}
  {2009})}\BibitemShut {NoStop}%
\bibitem [{\citenamefont {Giannozzi}\ \emph {et~al.}(2017)\citenamefont
  {Giannozzi}, \citenamefont {Andreussi}, \citenamefont {Brumme}, \citenamefont
  {Bunau}, \citenamefont {Nardelli}, \citenamefont {Calandra}, \citenamefont
  {Car}, \citenamefont {Cavazzoni}, \citenamefont {Ceresoli}, \citenamefont
  {Cococcioni}, \citenamefont {Colonna}, \citenamefont {Carnimeo},
  \citenamefont {Corso}, \citenamefont {de~Gironcoli}, \citenamefont {Delugas},
  \citenamefont {DiStasio}, \citenamefont {Ferretti}, \citenamefont {Floris},
  \citenamefont {Fratesi}, \citenamefont {Fugallo}, \citenamefont {Gebauer},
  \citenamefont {Gerstmann}, \citenamefont {Giustino}, \citenamefont {Gorni},
  \citenamefont {Jia}, \citenamefont {Kawamura}, \citenamefont {Ko},
  \citenamefont {Kokalj}, \citenamefont {Küçükbenli}, \citenamefont
  {Lazzeri}, \citenamefont {Marsili}, \citenamefont {Marzari}, \citenamefont
  {Mauri}, \citenamefont {Nguyen}, \citenamefont {Nguyen}, \citenamefont {de-la
  Roza}, \citenamefont {Paulatto}, \citenamefont {Poncé}, \citenamefont
  {Rocca}, \citenamefont {Sabatini}, \citenamefont {Santra}, \citenamefont
  {Schlipf}, \citenamefont {Seitsonen}, \citenamefont {Smogunov}, \citenamefont
  {Timrov}, \citenamefont {Thonhauser}, \citenamefont {Umari}, \citenamefont
  {Vast}, \citenamefont {Wu},\ and\ \citenamefont {Baroni}}]{Giannozzi_2017}%
  \BibitemOpen
  \bibfield  {author} {\bibinfo {author} {\bibfnamefont {P.}~\bibnamefont
  {Giannozzi}}, \bibinfo {author} {\bibfnamefont {O.}~\bibnamefont
  {Andreussi}}, \bibinfo {author} {\bibfnamefont {T.}~\bibnamefont {Brumme}},
  \bibinfo {author} {\bibfnamefont {O.}~\bibnamefont {Bunau}}, \bibinfo
  {author} {\bibfnamefont {M.~B.}\ \bibnamefont {Nardelli}}, \bibinfo {author}
  {\bibfnamefont {M.}~\bibnamefont {Calandra}}, \bibinfo {author}
  {\bibfnamefont {R.}~\bibnamefont {Car}}, \bibinfo {author} {\bibfnamefont
  {C.}~\bibnamefont {Cavazzoni}}, \bibinfo {author} {\bibfnamefont
  {D.}~\bibnamefont {Ceresoli}}, \bibinfo {author} {\bibfnamefont
  {M.}~\bibnamefont {Cococcioni}}, \bibinfo {author} {\bibfnamefont
  {N.}~\bibnamefont {Colonna}}, \bibinfo {author} {\bibfnamefont
  {I.}~\bibnamefont {Carnimeo}}, \bibinfo {author} {\bibfnamefont {A.~D.}\
  \bibnamefont {Corso}}, \bibinfo {author} {\bibfnamefont {S.}~\bibnamefont
  {de~Gironcoli}}, \bibinfo {author} {\bibfnamefont {P.}~\bibnamefont
  {Delugas}}, \bibinfo {author} {\bibfnamefont {R.~A.}\ \bibnamefont
  {DiStasio}}, \bibinfo {author} {\bibfnamefont {A.}~\bibnamefont {Ferretti}},
  \bibinfo {author} {\bibfnamefont {A.}~\bibnamefont {Floris}}, \bibinfo
  {author} {\bibfnamefont {G.}~\bibnamefont {Fratesi}}, \bibinfo {author}
  {\bibfnamefont {G.}~\bibnamefont {Fugallo}}, \bibinfo {author} {\bibfnamefont
  {R.}~\bibnamefont {Gebauer}}, \bibinfo {author} {\bibfnamefont
  {U.}~\bibnamefont {Gerstmann}}, \bibinfo {author} {\bibfnamefont
  {F.}~\bibnamefont {Giustino}}, \bibinfo {author} {\bibfnamefont
  {T.}~\bibnamefont {Gorni}}, \bibinfo {author} {\bibfnamefont
  {J.}~\bibnamefont {Jia}}, \bibinfo {author} {\bibfnamefont {M.}~\bibnamefont
  {Kawamura}}, \bibinfo {author} {\bibfnamefont {H.-Y.}\ \bibnamefont {Ko}},
  \bibinfo {author} {\bibfnamefont {A.}~\bibnamefont {Kokalj}}, \bibinfo
  {author} {\bibfnamefont {E.}~\bibnamefont {Küçükbenli}}, \bibinfo {author}
  {\bibfnamefont {M.}~\bibnamefont {Lazzeri}}, \bibinfo {author} {\bibfnamefont
  {M.}~\bibnamefont {Marsili}}, \bibinfo {author} {\bibfnamefont
  {N.}~\bibnamefont {Marzari}}, \bibinfo {author} {\bibfnamefont
  {F.}~\bibnamefont {Mauri}}, \bibinfo {author} {\bibfnamefont {N.~L.}\
  \bibnamefont {Nguyen}}, \bibinfo {author} {\bibfnamefont {H.-V.}\
  \bibnamefont {Nguyen}}, \bibinfo {author} {\bibfnamefont {A.~O.}\
  \bibnamefont {de-la Roza}}, \bibinfo {author} {\bibfnamefont
  {L.}~\bibnamefont {Paulatto}}, \bibinfo {author} {\bibfnamefont
  {S.}~\bibnamefont {Poncé}}, \bibinfo {author} {\bibfnamefont
  {D.}~\bibnamefont {Rocca}}, \bibinfo {author} {\bibfnamefont
  {R.}~\bibnamefont {Sabatini}}, \bibinfo {author} {\bibfnamefont
  {B.}~\bibnamefont {Santra}}, \bibinfo {author} {\bibfnamefont
  {M.}~\bibnamefont {Schlipf}}, \bibinfo {author} {\bibfnamefont {A.~P.}\
  \bibnamefont {Seitsonen}}, \bibinfo {author} {\bibfnamefont {A.}~\bibnamefont
  {Smogunov}}, \bibinfo {author} {\bibfnamefont {I.}~\bibnamefont {Timrov}},
  \bibinfo {author} {\bibfnamefont {T.}~\bibnamefont {Thonhauser}}, \bibinfo
  {author} {\bibfnamefont {P.}~\bibnamefont {Umari}}, \bibinfo {author}
  {\bibfnamefont {N.}~\bibnamefont {Vast}}, \bibinfo {author} {\bibfnamefont
  {X.}~\bibnamefont {Wu}}, \ and\ \bibinfo {author} {\bibfnamefont
  {S.}~\bibnamefont {Baroni}},\ }\href {\doibase 10.1088/1361-648X/aa8f79}
  {\bibfield  {journal} {\bibinfo  {journal} {Journal of Physics: Condensed
  Matter}\ }\textbf {\bibinfo {volume} {29}},\ \bibinfo {pages} {465901}
  (\bibinfo {year} {2017})}\BibitemShut {NoStop}%
\bibitem [{\citenamefont {{Davide Ceresoli}}(2023)}]{gipaw-website}%
  \BibitemOpen
  \bibfield  {author} {\bibinfo {author} {\bibnamefont {{Davide Ceresoli}}},\
  }\href@noop {} {\enquote {\bibinfo {title} {{GIPAW Code}},}\ } (\bibinfo
  {year} {2023}),\ \bibinfo {note}
  {\url{https://github.com/dceresoli/qe-gipaw}, Last accessed on
  2023-05-12}\BibitemShut {NoStop}%
\bibitem [{\citenamefont {Gorni}\ \emph {et~al.}(2018)\citenamefont {Gorni},
  \citenamefont {Timrov},\ and\ \citenamefont {Baroni}}]{Gorni:2018}%
  \BibitemOpen
  \bibfield  {author} {\bibinfo {author} {\bibfnamefont {T.}~\bibnamefont
  {Gorni}}, \bibinfo {author} {\bibfnamefont {I.}~\bibnamefont {Timrov}}, \
  and\ \bibinfo {author} {\bibfnamefont {S.}~\bibnamefont {Baroni}},\
  }\href@noop {} {\bibfield  {journal} {\bibinfo  {journal} {Eur. Phys. J. B}\
  }\textbf {\bibinfo {volume} {91}},\ \bibinfo {pages} {249} (\bibinfo {year}
  {2018})}\BibitemShut {NoStop}%
\bibitem [{\citenamefont {Kucukbenli}\ \emph {et~al.}(2014)\citenamefont
  {Kucukbenli}, \citenamefont {Monni}, \citenamefont {Adetunji}, \citenamefont
  {Ge}, \citenamefont {Adebayo}, \citenamefont {Marzari}, \citenamefont
  {{de~Gironcoli}},\ and\ \citenamefont {{Dal~Corso}}}]{Kucukbenli:2014}%
  \BibitemOpen
  \bibfield  {author} {\bibinfo {author} {\bibfnamefont {E.}~\bibnamefont
  {Kucukbenli}}, \bibinfo {author} {\bibfnamefont {M.}~\bibnamefont {Monni}},
  \bibinfo {author} {\bibfnamefont {B.}~\bibnamefont {Adetunji}}, \bibinfo
  {author} {\bibfnamefont {X.}~\bibnamefont {Ge}}, \bibinfo {author}
  {\bibfnamefont {G.}~\bibnamefont {Adebayo}}, \bibinfo {author} {\bibfnamefont
  {N.}~\bibnamefont {Marzari}}, \bibinfo {author} {\bibfnamefont
  {S.}~\bibnamefont {{de~Gironcoli}}}, \ and\ \bibinfo {author} {\bibfnamefont
  {A.}~\bibnamefont {{Dal~Corso}}},\ }\href@noop {} {\bibfield  {journal}
  {\bibinfo  {journal} {arXiv:1404.3015}\ } (\bibinfo {year}
  {2014})}\BibitemShut {NoStop}%
\bibitem [{\citenamefont {Kulik}\ and\ \citenamefont
  {Marzari}(2011)}]{Kulik:2011}%
  \BibitemOpen
  \bibfield  {author} {\bibinfo {author} {\bibfnamefont {H.}~\bibnamefont
  {Kulik}}\ and\ \bibinfo {author} {\bibfnamefont {N.}~\bibnamefont
  {Marzari}},\ }\href@noop {} {\bibfield  {journal} {\bibinfo  {journal} {J.
  Chem. Phys.}\ }\textbf {\bibinfo {volume} {134}},\ \bibinfo {pages} {094103}
  (\bibinfo {year} {2011})}\BibitemShut {NoStop}%
\bibitem [{\citenamefont {L\"owdin}(1950)}]{Lowdin:1950}%
  \BibitemOpen
  \bibfield  {author} {\bibinfo {author} {\bibfnamefont {P.-O.}\ \bibnamefont
  {L\"owdin}},\ }\href@noop {} {\bibfield  {journal} {\bibinfo  {journal} {J.
  Chem. Phys.}\ }\textbf {\bibinfo {volume} {18}},\ \bibinfo {pages} {365}
  (\bibinfo {year} {1950})}\BibitemShut {NoStop}%
\bibitem [{Mat()}]{MaterialsCloud}%
  \BibitemOpen
  \href@noop {} {}\bibinfo {note} {{The SSSP library of the Materials Cloud:
  \url{https://www.materialscloud.org/discover/sssp/table/precision}}}\BibitemShut
  {NoStop}%
\bibitem [{\citenamefont {Bonf\`a}\ \emph {et~al.}()\citenamefont {Bonf\`a},
  \citenamefont {John~Onuorah}, \citenamefont {Lang}, \citenamefont {Timrov},
  \citenamefont {Monacelli}, \citenamefont {Wang}, \citenamefont {Sun},
  \citenamefont {Petracic}, \citenamefont {Pizzi}, \citenamefont {Marzari},
  \citenamefont {Blundell},\ and\ \citenamefont
  {De~Renzi}}]{MaterialsCloudArchive2023}%
  \BibitemOpen
  \bibfield  {author} {\bibinfo {author} {\bibfnamefont {P.}~\bibnamefont
  {Bonf\`a}}, \bibinfo {author} {\bibfnamefont {I.}~\bibnamefont
  {John~Onuorah}}, \bibinfo {author} {\bibfnamefont {F.}~\bibnamefont {Lang}},
  \bibinfo {author} {\bibfnamefont {I.}~\bibnamefont {Timrov}}, \bibinfo
  {author} {\bibfnamefont {L.}~\bibnamefont {Monacelli}}, \bibinfo {author}
  {\bibfnamefont {C.}~\bibnamefont {Wang}}, \bibinfo {author} {\bibfnamefont
  {X.}~\bibnamefont {Sun}}, \bibinfo {author} {\bibfnamefont {O.}~\bibnamefont
  {Petracic}}, \bibinfo {author} {\bibfnamefont {G.}~\bibnamefont {Pizzi}},
  \bibinfo {author} {\bibfnamefont {N.}~\bibnamefont {Marzari}}, \bibinfo
  {author} {\bibfnamefont {S.~J.}\ \bibnamefont {Blundell}}, \ and\ \bibinfo
  {author} {\bibfnamefont {R.}~\bibnamefont {De~Renzi}},\ }\href {\doibase
  10.24435/materialscloud:8s-qh} {\enquote {\bibinfo {title}
  {{Magnetostriction-driven muon localisation in an antiferromagnetic
  oxide}},}\ }\bibinfo {howpublished} {Materials Cloud Archive \textbf{2023.82}
  (2023), doi:~10.24435/materialscloud:8s-qh}\BibitemShut {NoStop}%
\bibitem [{\citenamefont {Mayer}(2002)}]{Mayer:2002}%
  \BibitemOpen
  \bibfield  {author} {\bibinfo {author} {\bibfnamefont {I.}~\bibnamefont
  {Mayer}},\ }\href@noop {} {\bibfield  {journal} {\bibinfo  {journal} {Int. J.
  Quant. Chem.}\ }\textbf {\bibinfo {volume} {90}},\ \bibinfo {pages} {63}
  (\bibinfo {year} {2002})}\BibitemShut {NoStop}%
\bibitem [{\citenamefont {Baldereschi}(1973)}]{PhysRevB.7.5212}%
  \BibitemOpen
  \bibfield  {author} {\bibinfo {author} {\bibfnamefont {A.}~\bibnamefont
  {Baldereschi}},\ }\href {\doibase 10.1103/PhysRevB.7.5212} {\bibfield
  {journal} {\bibinfo  {journal} {Phys. Rev. B}\ }\textbf {\bibinfo {volume}
  {7}},\ \bibinfo {pages} {5212} (\bibinfo {year} {1973})}\BibitemShut
  {NoStop}%
\bibitem [{\citenamefont {Prandini}\ \emph {et~al.}(2018)\citenamefont
  {Prandini}, \citenamefont {Marrazzo}, \citenamefont {Castelli}, \citenamefont
  {Mounet},\ and\ \citenamefont {Marzari}}]{prandini2018precision}%
  \BibitemOpen
  \bibfield  {author} {\bibinfo {author} {\bibfnamefont {G.}~\bibnamefont
  {Prandini}}, \bibinfo {author} {\bibfnamefont {A.}~\bibnamefont {Marrazzo}},
  \bibinfo {author} {\bibfnamefont {I.~E.}\ \bibnamefont {Castelli}}, \bibinfo
  {author} {\bibfnamefont {N.}~\bibnamefont {Mounet}}, \ and\ \bibinfo {author}
  {\bibfnamefont {N.}~\bibnamefont {Marzari}},\ }\href@noop {} {\bibfield
  {journal} {\bibinfo  {journal} {npj Computational Materials}\ }\textbf
  {\bibinfo {volume} {4}},\ \bibinfo {pages} {1} (\bibinfo {year}
  {2018})}\BibitemShut {NoStop}%
\bibitem [{\citenamefont {Pratt}(2000)}]{PRATT2000710}%
  \BibitemOpen
  \bibfield  {author} {\bibinfo {author} {\bibfnamefont {F.}~\bibnamefont
  {Pratt}},\ }\href {\doibase https://doi.org/10.1016/S0921-4526(00)00328-8}
  {\bibfield  {journal} {\bibinfo  {journal} {Physica B: Condensed Matter}\
  }\textbf {\bibinfo {volume} {289-290}},\ \bibinfo {pages} {710} (\bibinfo
  {year} {2000})}\BibitemShut {NoStop}%
\bibitem [{\citenamefont {Timrov}\ \emph {et~al.}(2020)\citenamefont {Timrov},
  \citenamefont {Aquilante}, \citenamefont {Binci}, \citenamefont
  {Cococcioni},\ and\ \citenamefont {Marzari}}]{Timrov:2020b}%
  \BibitemOpen
  \bibfield  {author} {\bibinfo {author} {\bibfnamefont {I.}~\bibnamefont
  {Timrov}}, \bibinfo {author} {\bibfnamefont {F.}~\bibnamefont {Aquilante}},
  \bibinfo {author} {\bibfnamefont {L.}~\bibnamefont {Binci}}, \bibinfo
  {author} {\bibfnamefont {M.}~\bibnamefont {Cococcioni}}, \ and\ \bibinfo
  {author} {\bibfnamefont {N.}~\bibnamefont {Marzari}},\ }\href@noop {}
  {\bibfield  {journal} {\bibinfo  {journal} {Phys. Rev. B}\ }\textbf {\bibinfo
  {volume} {102}},\ \bibinfo {pages} {235159} (\bibinfo {year}
  {2020})}\BibitemShut {NoStop}%
\bibitem [{\citenamefont {Timrov}\ \emph {et~al.}(2022)\citenamefont {Timrov},
  \citenamefont {Marzari},\ and\ \citenamefont {Cococcioni}}]{Timrov:2022}%
  \BibitemOpen
  \bibfield  {author} {\bibinfo {author} {\bibfnamefont {I.}~\bibnamefont
  {Timrov}}, \bibinfo {author} {\bibfnamefont {N.}~\bibnamefont {Marzari}}, \
  and\ \bibinfo {author} {\bibfnamefont {M.}~\bibnamefont {Cococcioni}},\
  }\href@noop {} {\bibfield  {journal} {\bibinfo  {journal} {Comput. Phys.
  Commun.}\ }\textbf {\bibinfo {volume} {279}},\ \bibinfo {pages} {108455}
  (\bibinfo {year} {2022})}\BibitemShut {NoStop}%
\bibitem [{\citenamefont {Wang}\ \emph {et~al.}(2016)\citenamefont {Wang},
  \citenamefont {Chen},\ and\ \citenamefont {Jiang}}]{Wang:2016}%
  \BibitemOpen
  \bibfield  {author} {\bibinfo {author} {\bibfnamefont {Y.-C.}\ \bibnamefont
  {Wang}}, \bibinfo {author} {\bibfnamefont {Z.-H.}\ \bibnamefont {Chen}}, \
  and\ \bibinfo {author} {\bibfnamefont {H.}~\bibnamefont {Jiang}},\
  }\href@noop {} {\bibfield  {journal} {\bibinfo  {journal} {J. Chem. Phys.}\
  }\textbf {\bibinfo {volume} {144}},\ \bibinfo {pages} {144106} (\bibinfo
  {year} {2016})}\BibitemShut {NoStop}%
\bibitem [{\citenamefont {Tablero}(2008)}]{Tablero:2008}%
  \BibitemOpen
  \bibfield  {author} {\bibinfo {author} {\bibfnamefont {C.}~\bibnamefont
  {Tablero}},\ }\href@noop {} {\bibfield  {journal} {\bibinfo  {journal} {J.
  Phys.: Condens. Matter}\ }\textbf {\bibinfo {volume} {20}},\ \bibinfo {pages}
  {325205} (\bibinfo {year} {2008})}\BibitemShut {NoStop}%
\bibitem [{\citenamefont {Amadon}\ \emph {et~al.}(2008)\citenamefont {Amadon},
  \citenamefont {Jollet},\ and\ \citenamefont {Torrent}}]{Amadon:2008}%
  \BibitemOpen
  \bibfield  {author} {\bibinfo {author} {\bibfnamefont {B.}~\bibnamefont
  {Amadon}}, \bibinfo {author} {\bibfnamefont {F.}~\bibnamefont {Jollet}}, \
  and\ \bibinfo {author} {\bibfnamefont {M.}~\bibnamefont {Torrent}},\
  }\href@noop {} {\bibfield  {journal} {\bibinfo  {journal} {Phys. Rev. B}\
  }\textbf {\bibinfo {volume} {77}},\ \bibinfo {pages} {155104} (\bibinfo
  {year} {2008})}\BibitemShut {NoStop}%
\bibitem [{\citenamefont {Nawa}\ \emph {et~al.}(2018)\citenamefont {Nawa},
  \citenamefont {Akiyama}, \citenamefont {Ito}, \citenamefont {Nakamura},
  \citenamefont {Oguchi},\ and\ \citenamefont {Weinert}}]{Nawa:2018}%
  \BibitemOpen
  \bibfield  {author} {\bibinfo {author} {\bibfnamefont {K.}~\bibnamefont
  {Nawa}}, \bibinfo {author} {\bibfnamefont {T.}~\bibnamefont {Akiyama}},
  \bibinfo {author} {\bibfnamefont {T.}~\bibnamefont {Ito}}, \bibinfo {author}
  {\bibfnamefont {K.}~\bibnamefont {Nakamura}}, \bibinfo {author}
  {\bibfnamefont {T.}~\bibnamefont {Oguchi}}, \ and\ \bibinfo {author}
  {\bibfnamefont {M.}~\bibnamefont {Weinert}},\ }\href@noop {} {\bibfield
  {journal} {\bibinfo  {journal} {Phys. Rev. B}\ }\textbf {\bibinfo {volume}
  {97}},\ \bibinfo {pages} {035117} (\bibinfo {year} {2018})}\BibitemShut
  {NoStop}%
\end{thebibliography}%

\end{document}